\documentclass [12pt]{article}
\usepackage{amsmath,amssymb,cite}

\usepackage{tabularx}
\usepackage[english]{babel}
\usepackage[dvips]{graphicx}
\usepackage{indentfirst}
\usepackage{epsfig}
\begin{document}

\title{The One-Plaquette Model Limit of NC Gauge Theory in $2$D }

\author{Badis Ydri\footnote{Current Address : Institut fur Physik, Mathematisch-Naturwissenschaftliche Fakultat I, Humboldt-universitat zu Berlin, D-12489 Berlin-Germany.}\\
Department of Physics, Faculty of Science,\\
 Badji Mokhtar-Annaba University,
 Annaba, Algeria.}

\maketitle

\begin{abstract}
It is found that  noncommutative $U(1)$ gauge field on the fuzzy sphere  ${\bf S}^2_N$ is equivalent  in the quantum theory to a commutative $2-$dimensional $U(N)$ gauge field on a lattice with two plaquettes in the axial gauge $A_1=0$. This quantum equivalence holds in the fuzzy sphere-weak coupling phase in the limit of infinite mass of the scalar normal component of the gauge field.  The doubling of plaquettes is a natural consequence of the model and it is reminiscent of the usual doubling of points in Connes standard model. In the continuum large $N$ limit the plaquette variable $W$ approaches the identity ${\bf 1}_{2N}$ and as a consequence the model reduces to a simple matrix model which can be easily solved. We compute the one-plaquette critical point and show that it agrees with the observed value $\bar{\alpha}_*=3.35$. We compute  the quantum effective potential and the specific heat  for $U(1)$ gauge field on the fuzzy sphere
$S^2_{N}$ in the $1/N$ expansion using this one-plaquette model. In particular the specific heat per one degree of freedom was found to be equal to $1$ in the fuzzy sphere-weak coupling phase of the gauge field which agrees with the observed value $1$ seen in Monte Carlo simulation. This value of $1$ comes precisely because we have $2$ plaquettes approximating the NC $U(1)$ gauge  field on the fuzzy sphere.

\end{abstract}
\tableofcontents

\section{Introduction}

Quantum noncommutative ( NC ) gauge theory is essentially unknown beyond one-loop \cite{szabor}. In the one-loop approximation of the quantum theory  we know for example that gauge models on the Moyal-Weyl spaces are renormalizable \cite{martin}. These models were also shown to behave in a variety of novel ways as compared with their commutative counterparts. There are potential problems with unitarity and causality when time is noncommuting, and most notably we mention the notorious UV-IR mixing phenomena which is a generic property of all quantum field theories on Moyal-Weyl spaces and on noncommutative spaces in general \cite{szabor,min}. However a non-perturbative study of pure two dimensional noncommutative gauge theory was then performed in \cite{jun-wolf}.  For scalar field theory  on the Moyal-Weyl space some interesting non-perturbative results using theoretical and Monte Carlo methods were obtained for example in \cite{monte1}. An extensive list of references on these issues can be found in \cite{szabor} and also in \cite{others1}

The fuzzy sphere ( and any fuzzy space in general ) is designed for the study of gauge theories
in the non-perturbative regime using Monte-Carlo simulations. This is the point of view advocated in \cite{thesis}. See also \cite{madore,ars, iso} for quantum gravity, string theory or other different motivations. These fuzzy spaces consist in replacing continous manifolds
by matrix algebras and as a consequence the resulting field theory will only have  a finite number of
degrees of freedom. The claim is that this method has the advantage 
-in contrast with lattice- of preserving all continous symmetries of the original action at least at the classical level. This proposal was applied to the scalar ${\phi}^4$ model in \cite{xavier} and to the $U(1)$ gauge field in \cite{ref1} with very interesting non-perturbative results. Quantum field theory on fuzzy spaces was also studied perturbatively quite extensively. See for example \cite{perturbation,badis2,S2S2}. For some other non-perturbative ( theoretical or Monte Carlo ) treatement of these field theories see \cite{nonperturbative,steinackers2}.

Another motivation for considering the fuzzy sphere is the following. The Moyal-Weyl NC space is an infinite dimensional matrix model and not a continuum manifold and as a consequence it should be regularized by a finite dimensional matrix model. In $2$ dimensions the most natural candidate is the fuzzy sphere $S^2_N$ which is a finite dimensional matrix model which reduces to the NC plane in some appropriate large $N$ flattening limit. This limit was investigated quantum mechanically in \cite{badis2,badis1}. In $4-$dimensions we should instead consider Cartesian products of the fuzzy sphere $S^2_N$ \cite{S2S2}, fuzzy ${\bf CP}^2_N$ \cite{CP2} or fuzzy ${\bf S}^4$ \cite{S4}. It is fair to mention here that an alternative way of regularizing gauge theories  on the Moyal-Weyl NC space is based on the matrix model formulation of the twisted Eguchi-Kawai model. See for example \cite{kawai,szabor1,last30}.

The goal of this article and others \cite{ref1,ref} is to find the phase structure ( i.e map the different regions of the phase diagram ) of noncommutative  $U(1)$ gauge theories in $2$ dimensions on the fuzzy sphere ${\bf S}^2_N$.  
 We consider the fuzzy sphere since it is the most suited two dimensional space for numerical simulation because of the obvious fact that it is a well defined object. 

There are three phases of  $U(1)$ gauge theory on ${\bf S}^2_N$. In the matrix phase the fuzzy sphere vacuum collapses under quantum fluctuations and we have no underlying sphere in the continuum large $N$ limit. This phenomena was first observed in Monte Carlo simulation in \cite{nishimura} and then in \cite{ref1}. In \cite{ref} it was shown that the fuzzy sphere vacuum becomes more stable as the
mass of the scalar normal component of the gauge field increases. Hence this vacuum becomes completely stable when  this normal scalar field is projected out from the model. This is confirmed in Monte Carlo simulation in \cite{ref1}. 

In the other phase, the so-called fuzzy sphere phase, there are in fact two distinct regions in the phase diagram corresponding to the weak and strong coupling phases of the gauge field. The boundary between these two regions is demarcated by the usual third order one-plaquette phase transition \cite{gross}. This is precisely what we observe in our Monte Carlo simulation of the model with a very large mass of the normal scalar field \cite{ref1}. This result indicates that quantum noncommutative gauge theory is essentially equivalent ( at least in this fuzzy sphere phase ) to ( some ) commutative gauge theory not necessarily of the same rank. This prediction goes also in line with the powerful classical concept of Morita equivalence between noncommutative and commutative gauge theories on the torus \cite{szabor,szabor1}.

In this paper we will give a theoretical proof that quantum noncommutative gauge theory is equivalent to quantum commutative gauge theory in the fuzzy sphere-weak coupling phase in the limit of infinite mass of the normal scalar component of the gauge field. More precisely we will show that  the partition function of  a $U(1)$ gauge field on the fuzzy sphere  ${\bf S}^2_N$ is proportional to the partition function of  a {\it generalized} $2-$dimensional $U(N)$ gauge theory in the axial gauge  $A_1=0$  on a lattice with two plaquettes. This doubling of plaquettes is reminiscent of the usual doubling of points in Connes standard model \cite{connes}. This construction is based on the original fuzzy one-plaquette model due to \cite{peter1}. 

However in the present article we will show that in order to maintain gauge invariance and obtain sensible answers we will need to introduce two different $U(1)$ gauge fields on the fuzzy sphere which will only coincide in the continuum large $N$ limit. This doubling of fields is not related to the above doubling of plaquettes since it disappears in the continuum limit where the path integral is dominated by the configuration in which the two $U(1)$ gauge fields are equal. Furthermore we will need in the present article to  write down two different one-plaquette actions on the fuzzy sphere. Linear and quadratic terms in the plaquette variable $W$ are in fact needed in order to have convergence of the path integral. We will show explicitly the classical continuum limit of these one-plaquette actions. 

Quantum mechanically since the plaquette variable $W$ is small in the sense we will explain we can show that the model in the large $N$ limit will reduce to a simple matrix model and as a consequence can be easily solved. We compute the critical point and show that it agrees with the observed value. We will also compute the quantum effective potential for $U(1)$ gauge field on the fuzzy sphere
$S^2_{N}$ in the $1/N$ expansion using this one-plaquette model. This is in contrast with the calculation  of the effective potential in the limit $N\to\infty$ in the one-loop approximation done in \cite{ref}.  The difference between the two cases lies in the quantum logarithmic potential which is in absolute value larger by a factor of $4$ in the $1/N$ expansion as compared to the one-loop theory. We will discuss the implication of this to the critical point and possible interpretation of this result.  We will also compute the specific heat and find it equal to $1$ in the fuzzy sphere-weak coupling phase of the gauge field which agrees with the observed value $1$ seen in Monte Carlo simulation. The value $1$ comes precisely because we have two plaquettes which approximate the noncommutative $U(1)$ gauge field on the fuzzy sphere.

This paper is organized as follows. In section $2$ we will  briefly comment on the classical Morita equivalence between noncommutative gauge theories and commutative gauge theories on the torus. In section $3$ we will rederive the one-loop result of \cite{ref} using an RG method. Thus we will explicitly establish gauge invariance of the ${\bf S}^2_N$-to-matrix critical point. Section $4$ contains the main original results of this article discussed in the previous three paragraphs. In section $5$ we conclude with a summary and some general remarks.

\section{The NC torus and Morita equivalence}

The strongest argument concerning the equivalence between  classical noncommutative gauge theories and classical commutative gauge theories comes from considerations involving the noncommutative torus and Morita equivalence. In this section we will briefly review this result following the notations of \cite{szabor,szabor1}. 

Any $U(N)$ gauge model on the noncommutative torus $T^2_{\theta}$ with a non-zero magnetic flux $Q$ can be shown to be Morita equivalent to a $U(N_0)$ gauge model on the noncommutative torus $T^2_{{\theta}^{'}}$ with zero magnetic flux. 
The noncommutativity parameter ${\Theta}^{'}=2\pi {\theta}^{'}det {\Sigma}^{'}$ is given in terms of  $ {\Theta}=2\pi {\theta}det {\Sigma}$ by the  equation 
\begin{eqnarray}
{\Theta}^{'}=\frac{a\Theta -b}{m\Theta +l}.\label{ab}
\end{eqnarray}
The {\it integer} $l$ which is the ratio $l=N/N_0$ is the dimension of the irreducible representation of the Weyl-'t Hooft algebra 
\begin{eqnarray}
{\Gamma}_1{\Gamma}_2=e^{\frac{2 \pi}{N}i Q_{12}}{\Gamma}_2{\Gamma}_1 
\end{eqnarray}
found in the non-trivial solution 
\begin{eqnarray}
{\Omega}_a(x)=e^{i{\alpha}_{ai}x^i}\otimes{\Gamma}_a 
\end{eqnarray}
of the model 
\begin{eqnarray}
S_{YM}=-\frac{1}{4g^2}\int d^2x tr_N(F_{ij}-f_{ij})_*^2.
\end{eqnarray}
${\Gamma}_a$ are constant $SU(N)$ matrices while  ${\alpha}_{ai}$ is a $2 \times 2$ real matrix  which represents the $U(1)$ factor of the $U(N)$ group.

The $N \times N$ star-unitary transition functions ${\Omega}_a$ are {\it global large } gauge transformations whereas $f_{ij}$ is a constant curvature on $T^2_{{\theta}}$ which is equal to the curvature of the  $U(1)$ background gauge field $a_i$ given by  
\begin{eqnarray}
a_i=-\frac{1}{2}D_{ij}x^j\otimes {\bf 1}_N~,~D=2{\alpha}^T\frac{1}{\Sigma -\theta {\alpha}^T}. 
\end{eqnarray}
In above we have $q/N=m/l$ where $q=Q_{12}$ is the component of the antisymmetric matrix $Q$ of the non-abelian $SU(N)$ 't Hooft flux across the different non-contractible $2-$cycles of the noncommutative torus. By construction $q$  is quantized, i.e $q\in Z$. Furthermore $m$ is defined by $m=\frac{q}{x}$, $x={\rm gcd}(q,N)$ where `gcd'' stands for the great common divisor. Since ${l}$ and $m$ are relatively prime there exists two integers ${a}$ and $b$ such that $al+bm=1$. These are the same integers $a$ and $b$ which appear in (\ref{ab}).

The period matrix ${\Sigma}^{'}$ of the dual torus $T^2_{{\theta}^{'}}$ is related to the period matrix ${\Sigma}$ of the torus $T^2_{{\theta}}$ by 
\begin{eqnarray}
{\Sigma}^{'}=-(\Theta m +l)\Sigma.
\end{eqnarray}
The dual metric is therefore ${\eta}^{'}={\Sigma}^{'}{\Sigma}^{'T}=(\Theta m +l)^2 {\bf 1}$ which is to be compared with the original metric ${\eta}={\Sigma}{\Sigma}^{T}={\bf 1}$. 

The dual action is by the very definition of Morita equivalence equal to the $U(N)$ gauge action $S_{YM}$ on $T^2_{{\theta}}$, viz
\begin{eqnarray}
S_{YM}&=&-\frac{1}{4g^{'2}}\int d^2x^{'} tr_{N_0}({\cal F}_{ij}^{'}(x^{'}))_*^2\label{3}
\end{eqnarray}
where $g^{'2}$ is given in terms of $g^2$ by the equation 
\begin{eqnarray}
g^{'2}=g^2\frac{N_0}{N}(\Theta m +l)^2.
\end{eqnarray}
In other words $S_{YM}$ can also be interpreted as a $U(N_0)$ gauge action on  $T^2_{{\theta}^{'}}$. It is understood that the star product here is the one associated with the parameter ${\theta}^{'}$. It is the $U(1)$ background gauge field $a_i$  which is used to twist the boundary conditions on the $U(N)$ gauge field and hence obtain a non-trivial field configuration. Indeed the curvature $f_{12}$ of the vacuum gauge configuration $a_i$ on $T^2_{\theta}$ is related to the $SU(N)$  't Hooft magnetic flux $q$ by the equation 
 \begin{eqnarray}
f_{12}=\frac{1}{det \Sigma}\frac{2\pi q}{N+q\Theta}.
\end{eqnarray}
If we turn this equivalence upside down then we can obtain a correspondence between a $U(1)$ gauge model on a (finite dimensional) fuzzy torus $T^2_J$ and an ordinary $U(N)$ gauge model on $T^2$. In particular we remark that if we set $\Theta =0$, $\Sigma =1$ and $N_0=1$ in the above equations then $l=N$ and $m=q$ and we will have an ordinary $U(N)$ on a square torus $T^2$ with non-zero magnetic flux $q=\frac{Nf_{12}}{2\pi}$ and a coupling constant $g^2$. The dual torus in this case is also square since its period matrix is given by ${\Sigma}^{'}=-N.{\bf 1}$ whereas its noncommutativity parameter becomes
 \begin{eqnarray}
{\Theta}^{'}=-\frac{b}{N}
\end{eqnarray}
The commutation relation of the NC torus  $T^2_{{\theta}^{'}}$ becomes therefore 
 \begin{eqnarray}
\hat{z}_2^{'}\hat{z}_1^{'}&=&\hat{z}_1^{'}\hat{z}_2^{'} ~~exp(-2\pi i \frac{b}{N}).
\end{eqnarray}
Since the noncommutativity parameter here is rational we know that this Lie algebra must have a finite-dimensional $N{\times}N$  representation which can be written down in terms of shift and clock matrices as usual. In other words  
\begin{eqnarray}
&&\hat{z}_1^{'}=V_{N}~,~\hat{z}_2^{'}=\big( W_{N}\big)^{b}.
\end{eqnarray}
$V_{N}$ and $W_{N}$ are the canonical $SU(N)$ clock and shift matrices which satisfy $V_{N}W_{N}=e^{\frac{2 \pi i}{N}}W_{N}V_{N}$. This is indeed a fuzzy torus, i.e  $T^2_{{\theta}^{'}}=T^2_N$. The coupling constant of the $U(1)$ model on $T^2_N$ is $g^{'2}=g^2N$.

\section{The Fuzzy Sphere ${\bf S}^2_N$}

Let $X_a$ , $a=1,2,3$, be three $N{\times}N$ hermitian matrices and let us consider the action
\begin{eqnarray}
S[X_a]=-\frac{N}{4}Tr[X_a,X_b]^2+\frac{iN\alpha}{3}{\epsilon}_{abc}Tr[X_a,X_b]X_c+\beta TrX_a^2+MTr(X_a^2)^2 \label{ac1}
\end{eqnarray}
This action is invariant under $U(N)$ unitary transformations $X_a{\longrightarrow}UX_aU^{+}$. The trace is normalized such that $Tr1=N$. $\alpha$, $\beta$ and $M$ are the parameters of the model. This action is bounded from below for all strictly positive values of $M$. For $M=0$ this model  is also symmetric under global translations $X_a{\longrightarrow}X_a+{x}_a{\bf 1}_N$ where $\vec{x}$ is any constant vector. We can fix this symmetry by choosing the matrices $X_a$ to be traceless.

The classical equations of motion read 
\begin{eqnarray}
{\cal J}_a[X]=N{\alpha}^2[X_b,iX_{ab}]+2\beta X_a+2M\{X_b^2,X_a\}{\equiv}0~,~{\alpha}^2X_{ab}=i[X_a,X_b]+\alpha {\epsilon}_{abc}X_c.\label{ac4}
\end{eqnarray}
Absolute minima of the action are explicitly given by the fuzzy sphere solutions 
\begin{eqnarray}
X_a={R}L_a{\otimes}{\bf 1}_n.\label{ac2}
\end{eqnarray}
$L_a$ are the generators of $SU(2)$ in the irreducible representation $\frac{L}{2}$. They satisfy $[L_a,L_b]=i{\epsilon}_{abc}L_c$, $L_a^2=\frac{L}{2}(\frac{L}{2}+1){\equiv}c_2$ and  they are of size $(L+1){\times}(L+1)$, viz $N=n(L+1)$. $R$ is the radius of the sphere given explicitly by the solution of the equation
\begin{eqnarray}
(1+\frac{2c_2M}{N})R^2-{\alpha}R+\frac{\beta}{N}=0.\label{qua}
\end{eqnarray}
In particular for $\beta=M=0$ we have the solution $R=\alpha$. If we insist that $R=\alpha$ then we will have the constraint
\begin{eqnarray}
\beta=-2M{\alpha}^2c_2=-\tilde{\alpha}^2m^2~,~\tilde{\alpha}=\alpha \sqrt{N}~,~m^2=\frac{2c_2M}{N}. \label{1.15}
\end{eqnarray}
In general we can  show that a solution of (\ref{qua}) exists if and only if $\beta$ is such that $
\beta{\leq}\frac{\tilde{\alpha}^2}{4(1+m^2)}$.
Explicitly we have $\tilde{R}=R\sqrt{N}=\tilde{R}(\tilde{\alpha},m^2)$ with
\begin{eqnarray}
\tilde{R}(\tilde{\alpha},m^2)=\frac{\tilde{\alpha}+\sqrt{\tilde{\alpha}^2-4(1+m^2)\beta}}{2(1+m^2)}.
\end{eqnarray}
In the following we will strictly work with the case $R=\alpha$. We expand around the solution (\ref{ac2}) by writing
\begin{eqnarray}
X_a={\alpha}(L_a+A_a).
\end{eqnarray}
${A}_a$ , $a=1,2,3$ are $N{\times}N$ hermitian matrices which admit the interpretation of being the components of a $U(n)$ gauge field on a fuzzy sphere of size $(L+1){\times}(L+1)$. To see this we introduce the curvature tensor by
\begin{eqnarray}
{\alpha}^2F_{ab}=i[X_a,X_b]+{\alpha}{\epsilon}_{abc}X_c{\Leftrightarrow}F_{ab}=\bigg(i{\cal L}_aA_b-i{\cal L}_bA_a+{\epsilon}_{abc}A_c+i[A_a,A_b]\bigg).
\end{eqnarray}
We also introduce the normal component of $\vec{A}$ by
\begin{eqnarray}
{\alpha}^2\Phi=\frac{X_a^2-{\alpha}^2c_2}{2\sqrt{c_2}}{\Leftrightarrow}{\Phi}=\frac{1}{2}\bigg(x_aA_a+A_ax_a+\frac{A_a^2}{\sqrt{c_2}}\bigg)\label{normal}
\end{eqnarray}
where ${\cal L}_a=[L_a,..]$ and $x_a=\frac{L_a}{\sqrt{c_2}}$ are the derivations and coordinate-operators on the fuzzy sphere ${\bf S}^2_L$. We can then check that the action $S[X_a]$ takes  the form
\begin{eqnarray}
S[A_a]=\frac{{\tilde \alpha}^4}{4N}TrF_{ab}^2-\frac{\tilde{\alpha}^4}{4N}{\epsilon}_{abc}Tr\big[F_{ab}A_c-\frac{i}{3}[A_a,A_b]A_c\bigg]+\frac{2{\tilde \alpha}^4m^2}{N}Tr{\Phi}^2-\frac{1}{6}\tilde{\alpha}^4c_2-\frac{1}{2}\tilde{\alpha}^4c_2m^2.\label{ck}
\end{eqnarray}
We note that a natural definition of the $U(n)$ gauge coupling constant is given by ${g}^2=\frac{1}{\tilde{\alpha}^4}$. Also we note that $S_0\equiv S[A_a=0]= -\frac{1}{6}\tilde{\alpha}^4c_2-\frac{1}{2}\tilde{\alpha}^4c_2m^2$. Finally we remark that there is no  linear term in $\Phi$. For completeness we will include a linear term in $\Phi$ as follows
\begin{eqnarray}
\hat{S}[A_a]&=&\frac{{\tilde \alpha}^4}{4N}TrF_{ab}^2-\frac{\tilde{\alpha}^4}{4N}{\epsilon}_{abc}Tr\big[F_{ab}A_c-\frac{i}{3}[A_a,A_b]A_c\bigg]+\frac{2{\tilde \alpha}^4m^2}{N}Tr({\Phi}-{\phi}_0)^2+S_0\nonumber\\
&=&\frac{{\tilde \alpha}^4}{4N}TrF_{ab}^2-\frac{\tilde{\alpha}^4}{4N}{\epsilon}_{abc}Tr\big[F_{ab}A_c-\frac{i}{3}[A_a,A_b]A_c\bigg]+\frac{2{\tilde \alpha}^4m^2}{N}Tr{\Phi}^2-\frac{4{\tilde \alpha}^4m^2}{N}{\phi}_0Tr{\Phi}\nonumber\\
&+&\hat{S}_0.\label{ck10}
\end{eqnarray}
$\hat{S}_0=S_0++2{\tilde \alpha}^4m^2{\phi}_0^2$. Now in the limit $m{\longrightarrow}\infty $ the field $\Phi$ will be equal to a constant given by 
${\phi}_0$.
\subsection{The effective action  from an RG method}
We are interested in the partition function
\begin{eqnarray}
Z=\int [dX_a] e^{-S[X_a]}\label{ck1}
\end{eqnarray}
For simplicity we consider $U(1)$ theory so that $N=L+1$ and the full fuzzy $U(1)$ symmetry is given by the gauge group $U(N)$. The treatement of $U(n)$ is identical. We will also confine our analysis to the case where the coupling constant $\beta$ is related to $m$ by (\ref{1.15}). 

We will fix the $U(N)$ symmetry by diagonalizing the third matrix $X_3$. This will clearly reduce the original $U(N)$ symmetry group to its maximal abelian subgroup $U(1)^N$.  Although this method is not manifestly $SU(2)-$covariant it is completely gauge invariant since it does not require any extra parameter to be introduced in the model unlike other gauge-fixing procedures. Thus we will choose a unitary matrix $U$ such that $U^{+}X_3U={\Lambda}_3$ is a diagonal matrix with eigenvalues ${\lambda}_A$, $A=1,N$. We will have the simultaneous rotations $U^{+}X_iU={\Lambda}_i$, $i=1,2$. As it turns out $X_3=U{\Lambda}_3U^{+}$ can also be thought of as a parametrization of the matrix $X_3$ in terms of its radial degrees of freedom encoded in ${\Lambda}_3$ and its angular degrees of freedom given by $U=e^{i\Theta}$. Indeed we can compute the following metric and measure
\begin{eqnarray}
&&Tr(dX_3)^2=\sum_{A}d{\lambda}_A^2+2\sum_{A<B}({\lambda}_A-{\lambda}_B)^2d{\Theta}_{AB}d{\Theta}_{AB}^{*}\nonumber\\
&&[dX_3]=\bigg(\prod_{A=1}d{\lambda}_A\bigg)\bigg(\prod_{A<B}({\lambda}_A-{\lambda}_B)^2\bigg)\bigg({\prod}_{A<B}d{\Theta}_{ab}d{\Theta}_{ab}^{*}\bigg).
\end{eqnarray}
The partition function becomes ( since the integration over the unitary matrix $U$ decouples )
\begin{eqnarray}
Z=\int [d{\Lambda}_i] \bigg(\prod_{A=1}d{\lambda}_A\bigg) e^{-S[{\Lambda}_i,{\lambda}_A]}.
\end{eqnarray}
where the action is now given by  
\begin{eqnarray}
&&S[{\Lambda}_i,{\lambda}_A]=S_N[{\Lambda}_a]-\sum_{A<B}\log({\lambda}_A-{\lambda}_B)^2\nonumber\\
&&S_N[{\Lambda}_a]=-\frac{N}{4}Tr_{N}[{\Lambda}_a,{\Lambda}_b]^2+\frac{iN{\alpha}}{3}{\epsilon}_{abc}Tr_N[{\Lambda}_a,{\Lambda}_b]{\Lambda}_c+\beta Tr_N{\Lambda}_a^2+MTr_N ({\Lambda}_a^2)^2.\label{24}
\end{eqnarray}
We are using now the new notation $Tr{\equiv}Tr_N$. We stress again the fact that this action is still symmetric under the abelian $U(1)^N$ transformation ${\Lambda}_i{\longrightarrow}V^{+}{\Lambda}_iV$ and ${\Lambda}_3{\longrightarrow}{\Lambda}_3$ where $V$ is given explicitly by
\begin{eqnarray}
V_{AB}={\rm e}^{i{\theta}_A}{\delta}_{AB}.\label{abelian}
\end{eqnarray}
Now we adopt the RG prescription of \cite{brezin} to find the quantum corrections of this action at one-loop. To this end we parametrize the $N{\times}N$ matrices ${\Lambda}_a$ in terms of $(N-1){\times}(N-1)$ matrices $D_a$, $(N-1)-$dimensional vectors $v_a$ and $1-$dimensional vectors ${\rho}_a$ as follows
\begin{eqnarray}
{\Lambda}_a=\bigg( \begin{array}{cc}
        D_a & v_a \\
    v_a^{*} & {\rho}_a 
\end{array}\bigg).
\end{eqnarray}
Since ${\Lambda}_3$ is diagonal we must have $v_3=0$ while ${\rho}_3={\lambda}_N$. This method consists in finding quantum corrections to the action coming from integrating out the $4(N-1)+3$ degrees of freedom $v_i$ and ${\rho}_a$ which we can  naturally think of as  fluctuations around a  background defined by the matrices $D_a$. Furthermore it is not difficult to argue that this method is also equivalent to the usual Wilson procedure of integrating out the top modes with spin $L=N-1$ from the theory. 

To see this more explicitly we write  $({\Lambda}_a)_{AB}=(D_a)_{AB}$, $({\Lambda}_i)_{A N}=v_i^{A}$, $({\Lambda}_i)_{N A}=(v_i^{A})^{*}$ and $({\Lambda}_a)_{NN}={\rho}_a$ where $A,B=1,...,N-1$. We check that the abelian transformations (\ref{abelian}) will act on $D_a$, $v_i$ and ${\rho}_a$ as follows 
\begin{eqnarray}
D_a{\longrightarrow}WD_aW^{+}~,~v_a{\longrightarrow}Wv_a~,~v_a^{+}{\longrightarrow}v_a^{+}W^{+},~{\rho}_a{\longrightarrow}{\rho}_a\label{residual}
\end{eqnarray}
where
\begin{eqnarray}
(W_a)_{AB}=e^{-i({\theta}_A-{\theta}_N)}{\delta}_{AB}=e^{i{\theta}_N}(V^{+})_{AB}.
\end{eqnarray}
Next we will denote the $(N-1)-$dimensional trace by $Tr_{N-1}$ and  compute
\begin{eqnarray}
Tr_{N}[{\Lambda}_a,{\Lambda}_b]^2&=&Tr_{N-1}[D_a,D_b]^2+v_i^{+}\big[-4D_iD_j+8D_jD_i-4D_a^2{\delta}_{ij}\big]v_j+O(3).\label{decomp1}
\end{eqnarray}
\begin{eqnarray}
i{\epsilon}_{abc}Tr_N[{\Lambda}_a,{\Lambda}_b]{\Lambda}_c=i{\epsilon}_{abc}Tr_{N-1}[D_a,D_b]D_c-6i{\epsilon}_{ij3}v_i^{+}D_3v_j+O(3).
\end{eqnarray}
\begin{eqnarray}
Tr_N{\Lambda}_a^2=Tr_{N-1}D_a^2+2v_i^{+}v_i      +{\rho}_a^2.
\end{eqnarray}
and
\begin{eqnarray}
Tr_N({\Lambda}_a^2)^2=Tr_{N-1}(D_a^2)^2+\sum_{a=1}^3{\rho}_a^4+v_i^{+}\big[2D_a^2{\delta}_{ij}+2D_iD_j\big]v_i+O(3).\label{last}
\end{eqnarray}
$O(3)$ stands for cubic or higher order terms. 
In one-loop approximation it is sufficient that one keeps only terms up to quadratic powers in the fluctuation fields which are identified here with the $v_i$ and ${\rho}_a$ degrees of freedom. However we have kept the quartic term $\sum_{a=1}^3{\rho}_a^4$ in equation (\ref{last}) for other purposes which will become clearer shortly. The action $S_N[{\Lambda}_a]$  is then given by
\begin{eqnarray}
S_N[{\Lambda}_a]=S_{N-1}[D_a]+v_i^{+}{\Omega}_{ij}v_j+\sum_{a=1}^3[\beta {\rho}_a^2+M{\rho}_a^4]+O(3). \label{action}
\end{eqnarray}
where clearly
\begin{eqnarray}
S_{N-1}[{D}_a]=-\frac{N}{4}Tr_{N-1}[{D}_a,{D}_b]^2+\frac{iN{\alpha}}{3}{\epsilon}_{abc}Tr_{N-1}[{D}_a,{D}_b]{D}_c+\beta Tr_{N-1}{D}_a^2+MTr_{N-1} ({D}_a^2)^2.\nonumber\\\label{foundit}
\end{eqnarray}
The operators ${\Omega}_{ij}$ are given explicitly by
\begin{eqnarray}
&&{\Omega}_{ij}=\bigg(2 \beta +(2M+N)D_a^2\bigg){\delta}_{ij}+(2M-N)D_iD_j-2i\tilde{\alpha}^2F_{ij}.\label{exp}
\end{eqnarray}
From equations (\ref{residual}) and (\ref{action}) it is quite clear that the  $(N-1)-$dimensional vectors $v_i$ play exactly the role of (bosonic) quark fields  moving in the background of a covariant  $U(1)^{N-1}$ gauge field $D_a$. On the other hand the logarithmic potential ( equation (\ref{24}) )  takes the form
\begin{eqnarray}
\sum_{A<B}\log({\lambda}_A-{\lambda}_B)^2=\sum_{A<B}\log(d_A-d_B)^2+\sum_{A=1}^{N-1}\log(d_A-{\rho}_3)^2.
\end{eqnarray}
By integrating out $v_i$, $v^{*}_i$ and ${\rho}_a$ we obtain the effective action $S_{N-1}^{\rm eff}[D_i,d_a]$ given by
\begin{eqnarray}
e^{-S_{N-1}^{\rm eff}[D_i,d_A]}&=&e^{-S_{N-1}[D_a]+\sum_{A<B}\log(d_A-d_B)^2}\int {\prod}_{i=1}^2[dv_{i}^{*}][dv_i]e^{-v_i^{+}{\Omega}_{ij}v_j}\int {\prod}_{i=1}^2[d{\rho}_i]~e^{-\beta {\rho}_i^2-M{\rho}_i^4}\nonumber\\
&{\times}&\int d{\rho}_3e^{-{\beta}{\rho}_3^2-M{\rho}_3^4+\sum_{A=1}^{N-1}\log(d_A-{\rho}_3)^2}.\label{path}
\end{eqnarray}
The effective action reads therefore
\begin{eqnarray}
S_{N-1}^{\rm eff}[D_i,d_A]&=&S_{N-1}[D_a]-\sum_{A<B}\log(d_A-d_B)^2-\log\bigg[\int d{\rho}e^{-{\beta}{\rho}^2-M{\rho}^4+\sum_{A=1}^{N-1}\log(d_A-{\rho})^2}\bigg]\nonumber\\
&+&Tr_2Tr_{N-1}\log\bigg(\big(2 \beta +(2M+N)D_a^2\big){\delta}_{ij}+(2M-N)D_iD_j-2i\tilde{\alpha}^2F_{ij}\bigg).\label{218}\nonumber\\
\end{eqnarray}
$Tr_2$ is the $2-$dimensional trace associated with the remaining $U(1)$ rotational symmetry of the two matrices $D_1$ and $D_2$ ( since $D_3$ is treated differently -diagonalized-in this approach) and $Tr_{N-1}$ is the usual trace over the matrices; here $D_1$ and $D_2$ are $(N-1){\times}(N-1)$ matrices. 

The sum of the first two terms in (\ref{218}) is nothing but the action (\ref{24}) with the replacement ${\Lambda}_a{\longrightarrow}D_a$, ${\lambda}_A{\longrightarrow}d_A$ and $Tr_{N}{\longrightarrow}Tr_{N-1}$.  Thus
\begin{eqnarray}
S_{N-1}^{\rm eff}[D_i,d_A]&=&S_{N-1}[D_i,d_A]-\log\bigg[\int d{\rho}e^{-{\beta}{\rho}^2-M{\rho}^4+\sum_{A=1}^{N-1}\log(d_A-{\rho})^2}\bigg]\nonumber\\
&+&Tr_2Tr_{N-1}\log\bigg(\big(2 \beta +(2M+N)D_a^2\big){\delta}_{ij}+(2M-N)D_iD_j-2i\tilde{\alpha}^2F_{ij}\bigg).\nonumber\\
\end{eqnarray}
This action has $U(1)^{N-1}$ gauge symmetry and a $U(1)$ rotational symmetry since the matrix $D_3$ is diagonal.  In the partition function the gauge symmetry can be easily enlarged to $U(N-1)$ by rotating the diagonal matrix ${D}_3$ ( back ) to a general form $C_3$ given by ${D}_3=UC_3U^{+}$ where $U$ is an $(N-1){\times}(N-1)$ unitary matrix.  We will have the simultaneous rotations ${D}_i=UC_iU^{+}$. The action $S_{N-1}[D_i,d_A]$ becomes given by (\ref{ac1}) with the replacement $X_a{\longrightarrow}C_a$ and where the trace is normalized such that $Tr1=N-1$. We write the above result in the following suggestive form

\begin{eqnarray}
{\delta}S_{N-1}^{\rm eff}[D_a]&=&-\log\bigg[\int d{\rho}e^{-{\beta}{\rho}^2-M{\rho}^4+\sum_{A=1}^{N-1}\log(d_A-{\rho})^2}\bigg]\nonumber\\
&+&Tr_2Tr_{N-1}\log\bigg(\big(2 \beta +(2M+N)D_a^2\big){\delta}_{ij}+(2M-N)D_iD_j-2i\tilde{\alpha}^2F_{ij}\bigg).\nonumber\\\label{effe1}
\end{eqnarray}
${\delta}S_{N-1}^{\rm eff}$ is precisely the one-loop contribution to the classical action (\ref{ac1}) coming from integrating out  from the model only one row and one column. In the large $N$ limit we can treat $N$ as a continuous variable and thus we can simply obtain the full one-loop contribution  to the classical action (\ref{ac1}) by integration over $N$ of the above result.
\subsection{The ${\bf S}^2_N$-to-Matrix phase transition}
We are interested in particular in verifying the stability of the fuzzy sphere ground state (\ref{ac2}) under quantum fluctuations. We consider therefore the background $D_a=\alpha {\phi}L_a$ where $L_a$ are the generators of $SU(2)$ in the irreducible representation $\frac{N-2}{2}$ which are of size $(N-1){\times}(N-1)$. ${\phi}$ is the field associated with the fluctuations of the radius $R=\alpha$. The classical potential from (\ref{foundit}) is given by 
\begin{eqnarray}
V[\phi]&=&\frac{N-2}{N+1}\bigg[2c_2\tilde{\alpha}^4\bigg(\frac{1}{4}{\phi}^4-\frac{1}{3}{\phi}^3\bigg)\bigg]+\frac{N-2}{N+1}\bigg[c_2\tilde{\alpha}^2{\beta}{\phi}^2\bigg]+\frac{N-2}{N+1}\frac{(N-1)^2-1}{N^2-1}\bigg[\frac{c_2^2\tilde{\alpha}^4M}{N}{\phi}^4\bigg]\nonumber\\
&=&\frac{N^2\tilde{\alpha}^4}{2}\bigg[\frac{1+m^2}{4}{\phi}^4-\frac{1}{3}{\phi}^3-\frac{m^2}{2}{\phi}^2\bigg]+O(N).
\end{eqnarray}
where we have used the relations $\tilde{\alpha}=\alpha \sqrt{N}$, $\beta=-\tilde{\alpha}^2m^2$ and $M=\frac{Nm^2}{2c_2}$. We have also kept terms of order $N^2$ only which dominates in the large $N$ limit.  Since $D_3= \alpha \phi  L_3$ and $(L_3)_{AB}=m_A{\delta}_{AB}$ where $m_A=\frac{N}{2}-A$ we have $d_A=\alpha \phi m_A$. Thus the last integral in (\ref{path}) becomes in the large $N$ limit
\begin{eqnarray}
e^{(2N-1)\log\alpha \phi}\int d{\rho}e^{\frac{\tilde{\alpha}^4m^2{\phi}^2}{N}{\rho}^2-\frac{m^2\tilde{\alpha}^4{\phi}^4}{2c_2N}{\rho}^4+\sum_{A=1}^{N-1}\log(\frac{N}{2}-{\rho}-A)^2}{\simeq}~c~ e^{(2N-1)\log\phi}.\label{1.33}
\end{eqnarray} 
The constant $c$ is independent of the field $\phi$. We also compute in the large $N$ limit
\begin{eqnarray}
2\beta +(2M+N)D_a^2&=&\tilde{\alpha}^2\big(-2m^2+\frac{N^2-2N}{N^2-1}(c_2+m^2){\phi}^2\big)\nonumber\\
&=&\tilde{\alpha}^2\bigg(\frac{N^2}{4}{\phi}^2-\frac{N}{2}{\phi}^2-2m^2+m^2{\phi}^2-\frac{2m^2}{N}{\phi}^2+O(\frac{1}{N^2})\bigg).\label{1.34}\end{eqnarray}
\begin{eqnarray}
(2M-N)D_iD_j&=&\tilde{\alpha}^2\frac{N^2-2N}{N^2-1}(m^2-c_2){\phi}^2x_ix_j\nonumber\\
&=&\tilde{\alpha}^2\bigg(-\frac{N^2}{4}+\frac{N}{2}+m^2-\frac{2m^2}{N}+O(\frac{1}{N^2})\bigg){\phi}^2x_ix_j~,~x_i=\frac{2L_i}{\sqrt{N^2-2N}}.\label{1.35}\nonumber\\
\end{eqnarray}
\begin{eqnarray}
-2i\tilde{\alpha}^2F_{ij}&=&-2i\tilde{\alpha}^2\sqrt{c_2}\sqrt{\frac{N^2-2N}{N^2-1}}(\phi -{\phi}^2){\epsilon}_{ij3}x_3\nonumber\\
&=&-i\tilde{\alpha}^2\bigg({N}-1-\frac{1}{2N}+O(\frac{1}{N^2})\bigg)(\phi -{\phi}^2){\epsilon}_{ij3}x_3~,~x_3=\frac{2L_3}{\sqrt{N^2-2N}}.\label{1.36}
\end{eqnarray}
The leading quantum contribution of the effective potential from (\ref{effe1}),(\ref{1.33}) and  (\ref{1.34})-(\ref{1.36})  is given by

\begin{eqnarray}
{\delta}V_{N-1}^{\rm eff}&=&-(2N-1)\log\phi +Tr_2Tr_{N-1}\log\bigg[\frac{\tilde{\alpha}^2N^2{\phi}^2}{4}\big({\delta}_{ij}-x_ix_j\big)\bigg]\nonumber\\
&=&(2N-3)\log\phi+{\rm a~constant~independent~of}~\phi.\label{all}
\end{eqnarray}
As we will check in appendix $A$ all higher order terms in equations (\ref{1.34})-(\ref{1.36}) will give vanishingly small quantum contributions in the large $N$ limit. Thus (\ref{all}) is the one-loop correction of the effective potential coming from integrating out one column and one row from the theory. The full one-loop correction coming from integrating out all columns and rows is obtained by a simple integration over $N$. We get
\begin{eqnarray}
V_{N-1}^{\rm eff}&=&(N^2-3N)\log\phi.\label{effe2}
\end{eqnarray}
By putting (\ref{effe1}) and (\ref{effe2}) together we get the effective potential

\begin{eqnarray}
V_{\rm eff}&=&\frac{N^2\tilde{\alpha}^4}{2}\bigg[\frac{1+m^2}{4}{\phi}^4-\frac{1}{3}{\phi}^3-\frac{m^2}{2}{\phi}^2\bigg]+N^2\log\phi +O(N).\label{emp}
\end{eqnarray}
Let us point out here that this potential was derived elsewhere using a completely different (much simpler) argument involving gauge fixing the original action (\ref{ac1}) and then computing the full one-loop effective action in the background field method. The argument in this article is however superior from the point of view that it ( manifestly) preserves gauge symmetry at all stages of the calculation since we are not fixing any gauge in the usual sense \cite{ref}. See also \cite{nishimura}.
 
It is not difficult to check that the corresponding equation of motion of the potential (\ref{emp}) admits two real solutions where we can identify the one with the least energy  with the actual radius of the sphere. This however is only true up to a certain value $\tilde{\alpha}_{*}$ of the coupling constant $\tilde{\alpha}$ where the quartic equation ceases to have any real solution and as a consequence the fuzzy sphere solution (\ref{ac2}) ceases to exist. In other words the potential $V_{\rm eff}$ below the value $\tilde{\alpha}_{*}$ of the coupling constant becomes unbounded and the fuzzy sphere  collapses. The critical value can be easily computed and one finds
\begin{eqnarray}
{\phi}_{*}=\frac{3}{8(1+m^2)}\bigg[1+\sqrt{1+\frac{32m^2(1+m^2)}{9}}\bigg].\label{pre}
\end{eqnarray}
and 
\begin{eqnarray}
\frac{1}{\tilde{\alpha}^4_{*}}=-\frac{1}{2}(1+m^2){\phi}_{*}^4+\frac{1}{2}{\phi}_{*}^3+\frac{m^2}{2}{\phi}_{*}^2.\label{pre1}
\end{eqnarray}
Extrapolating to large masses  we obtain the scaling behaviour 
\begin{eqnarray}
\tilde{\alpha}_{*}=\big[\frac{8}{m^2+\sqrt{2}-1}\big]^{\frac{1}{4}}. \label{pre2}
\end{eqnarray}
In other words the phase transition happens each time at a smaller value of the coupling constant $\tilde{\alpha}$ and thus the fuzzy sphere is more stable. This one-loop result is compared to the non-perturbative result coming from  the Monte Carlo simulation of the model (\ref{ac1}) with the constraint (\ref{1.15}) in figure $1$. As one can immediately see there is an excellent agreement. In this sense the one-loop result for the $U(1)$ is exact. Let us finally report that this phase transition was also observed in $4$ dimensions on ${\bf S}^2_L\times{\bf S}^2_L$. See the first reference of \cite{S2S2}.

\begin{figure}[h]
\begin{center}
\includegraphics[width=7cm,angle=-90]{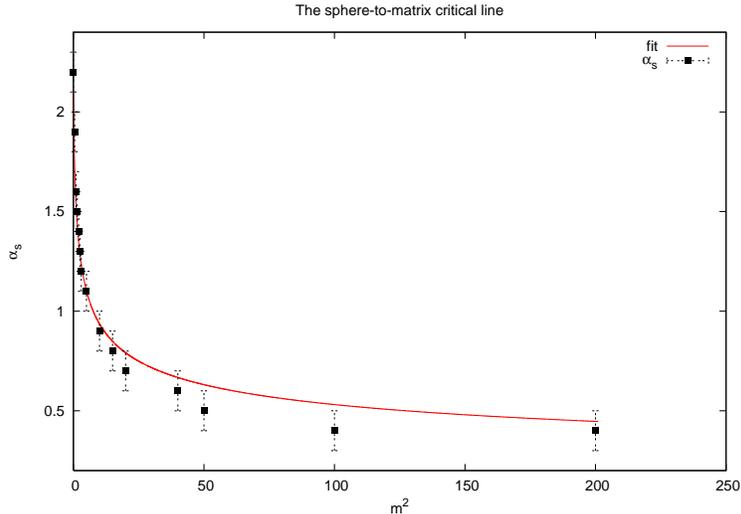}
\caption{{The phase diagram of the ${\bf S}^2_N-$to-matrix phase transition. The fuzzy sphere phase is above the solid line while the matrix phase is below it. In this figure ${\alpha}_s$ is the Monte Carlo measurement of the critical value $\tilde{\alpha}_{*}$.}}
\end{center}
\end{figure}

\section{The {\it small} one-plaquette model limit on ${\bf S}^2_N$}

We have found in the one-loop calculation as well as in numerical simulation that the presence of the normal field (\ref{normal}) is what causes the model to undergo the above first order phase transition from the fuzzy sphere to a matrix phase where the fuzzy sphere collapses under quantum fluctuation. At the level of perturbation theory of the gauge field $A_a$ this shows up in the form of a compact UV-IR mixing phenomena which goes to the usual singular UV-IR mixing on the NC plane in some appropriate planar limit of the sphere. In the large $m$ limit we also have shown that these two ( possibly related ) effects disappear \cite{ref}. 

As it turns out there is some signature  in Monte Carlo simulation of the model  (\ref{ac1}) with the constraint (\ref{1.15}) for the existence of another kind of phase transition which seems to be unrelated to the ${\bf S}^2_{N}$-to-matrix phase transition and which generically persists even in the large $m$ limit. This latter phase transition resembles very much the third order one-plaquette phase transition in ordinary two dimensional gauge theory. In particular the agreement in the weak regime between the simulation and the theory ( which we will present now ) is excellent. Let us also say that this transition starts to appear when the critical value $\tilde{\alpha}_{*}$ as $m$ increases becomes less than the value $\frac{3.35}{\sqrt{N}}$ and it becomes more pronounced as  $\tilde{\alpha}_{*}$ decreases further away from this value. The new phase transition thus occurs at 
\begin{eqnarray}
\tilde{\alpha}_{*}=\frac{3.35}{\sqrt{N}}.\label{336}
\end{eqnarray}
Our goal in this section is to give a detailed theoretical model which describes this transition. This construction is motivated by \cite{peter1,steinackers2}.

We start by making the observation that in the large $m{\longrightarrow}{\infty}$ limit we can set $\Phi =0$ as one can immediately see  from the action (\ref{ck}) and the partition function (\ref{ck1}).  Indeed we have  for $m{\longrightarrow}{\infty}$
\begin{eqnarray}
e^{-\frac{2m^2}{g^2N}Tr{\Phi}^2}=\bigg(\frac{g^2N\pi}{2m^2}\bigg)^{\frac{N^2}{2}}{\delta}(\Phi).
\end{eqnarray}
In other words the normal scalar field $\Phi$ becomes infinitely heavy ( $m$ is precisely its mass ) and thus decouples from the rest of the dynamics. Hence we can effectively impose the extra constraint $X_a^2={\alpha}^2c_2$ on the field $X_a$ in this limit $m{\longrightarrow}{\infty}$. In terms of $D_a=L_a+A_a=\frac{X_a}{\alpha}$ this constraint reads
\begin{eqnarray}
\Phi=\frac{D_a^2-c_2}{2\sqrt{2}}=\frac{1}{2}\{x_a,A_a\}+\frac{1}{2\sqrt{c_2}}A_a^2=0.
\end{eqnarray}
The action (\ref{ck}) ( if (\ref{1.15}) is also satisfied ) becomes 
\begin{eqnarray}
S&=&S_{YM}+S_{CS}+S_0+\frac{N^2}{2}\log\bigg(\frac{g^2N\pi}{2m^2}\bigg).
\end{eqnarray}
The Yang-Mills and
Chern-Simons-like actions are given respectively by
\begin{eqnarray}
S_{YM}&=&\frac{1}{4g^2N} TrF_{ab}^2\nonumber\\
~S_{CS}&=&-\frac{1}{6g^2N}Tr\big[\epsilon_{abc}F_{ab}D_c+D_a^2-c_2\big]=-\frac{1}{6g^2N}Tr\epsilon_{abc}F_{ab}D_c~{\rm when}~m{\longrightarrow}\infty .\label{YMCS}
\end{eqnarray}
In above ${g}^2=\frac{1}{\tilde{\alpha}^4}$ and $S_0=S[A_a=0]= -\frac{1}{6}\tilde{\alpha}^4c_2-\frac{1}{2}\tilde{\alpha}^4c_2m^2$
. 
In the continuum large $N{\longrightarrow}{\infty}$ limit the constraint  $X_a^2={\alpha}^2c_2$ becomes the usual requirement that the normal component of the gauge field on the sphere is zero, viz $\Phi=n_a.A_a=0$. Moreover the Chern-Simons-like action vanishes in this limit by this same condition $\Phi=n_a.A_a=0$ because it will involve the integral of a $3-$form over a $2-$dimensional manifold. Hence $S={S}_{YM}$ in the large $N{\longrightarrow}{\infty}$ limit provided we also impose the condition $\Phi=0$. In summary if we take the limit $m{\longrightarrow}{\infty}$ first and then we take the continuum limit $N{\longrightarrow}{\infty}$ we obtain a $U(1)$ action on the ordinary sphere, viz
\begin{eqnarray}
S-S_0&=&\frac{N^2}{2}\log\bigg(\frac{g^2N\pi}{2m^2}\bigg)+\frac{1}{4g^2} \int \frac{d{\Omega}}{4\pi}(i{\cal L}_aA_b-i{\cal L}_bA_a+{\epsilon}_{abc}A_c)^2.\label{aclim}
\end{eqnarray}
By construction the continuum gauge field ( which should be easily distinguished from its corresponding operator on the fuzzy sphere although we are using the same symbol $A_a$ for both quantities ) is strictly tangent. See appendix  $B$ for more detail.

If we study instead the action (\ref{ck10}) in the limit $m{\longrightarrow}{\infty}$ first and then in the continuum limit $N{\longrightarrow}{\infty}$ then we will find the action
\begin{eqnarray}
\hat{S}-{S}_0-{\tilde{\alpha}}^4{\phi}_0^2&=&\frac{1}{4g^2} \int \frac{d{\Omega}}{4\pi}(i{\cal L}_aA_b-i{\cal L}_bA_a+{\epsilon}_{abc}A_c)^2.
\end{eqnarray}
Again the continuum gauge field $A_a$ is strictly tangent. The only difference with the previous case is the extra piece ${\tilde{\alpha}}^4{\phi}_0^2$ which we pulled out from the action in the process of writing it only in terms of a tangent gauge field.

We now relate this action with the one-plaquette action. To this end we introduce the $2N{\times}2N$ idempotent 
\begin{eqnarray}
{\gamma}=\frac{1}{N}({\bf 1}_{2N}+2{\sigma}_aL_a)~,~{\gamma}^2=1
\end{eqnarray}
 where ${\sigma}_a$ are the usual Pauli matrices. It has eigenvalues $+1$ and $-1$ with multiplicities $N+1$ and $N-1$ respectively. We introduce the covariant derivative $D_a=L_a+A_a$ through a gauged idempotent ${\gamma}_D$ as follows
\begin{eqnarray}
&&{\gamma}_D=\hat{\gamma}\frac{1}{\sqrt{\hat{\gamma}^2}}\nonumber\\
&&\hat{\gamma}=\frac{1}{N}(1+2{\sigma}_aD_a)={\gamma}+\frac{2}{N}{\sigma}_aA_a~,~~\hat{\gamma}^2=1+\frac{8\sqrt{c_2}}{N^2}\Phi+\frac{2}{N^2}{\epsilon}_{abc}{\sigma}_cF_{ab}.\label{eq0}
\end{eqnarray}
Since we are interested in the large $m{\longrightarrow}{\infty}$ limit we may as well set $\Phi=0$ in above. 
Clearly ${\gamma}_D$ has the same spectrum as ${\gamma}$. Thus it is a continuous deformation of ${\gamma}$ in the sense that there exists a $U(2N)$ unitary transformation $U$ such that ${\gamma}_D=U{\gamma}U^{+}$. Furthermore if $U{\longrightarrow}UT$ where $T{\in}U(N+1)$ or $T{\in}U(N-1)$ then ${\gamma}{\longrightarrow}T{\gamma}T^{+}={\gamma}$ and as a consequence ${\gamma}_D{\longrightarrow}{\gamma}_D$. So ${\gamma}_D$ is an element of the $d_N-$Grassmannian manifold $U(2N)/U(N+1){\times}U(N-1)$. We compute the dimension $d_N$ as follows 
\begin{eqnarray}
d_N=4N^2-(N+1)^2-(N-1)^2=2N^2-2. \label{dimen}
\end{eqnarray}
This is exactly the correct number of degrees of freedom in a gauge theory on the sphere without normal scalar field or  with a normal scalar field frozen to some fixed value. The $2$ counts the zero modes which decouple because of commutators. 

The original $U(N)$ gauge symmetry acts on the covariant derivatives $D_a$ as 
\begin{eqnarray}
D_a^g=g{D}_ag^{+}~,~g{\in}U(N).\label{ori}
\end{eqnarray}
This symmetry  will be enlarged to the following $U(2N)$ symmetry. We introduce a tentative link variable $W$ ( a $2N{\times}2N$ unitary matrix ) by $W={\gamma}{\gamma}_D$. The extended  $U(2N)$ symmetry will then act on $W$ as follows $
W{\longrightarrow}VWV^{+}~,~V{\in}U(2N)$. 
It is clear that this transformation property of $W$ can only be obtained if we  impose the following transformation properties $\gamma{\longrightarrow}V\gamma V^{+}$ and ${\gamma}_D{\longrightarrow}V{\gamma}_DV^{+}$ on $\gamma$ and ${\gamma}_D$ respectively. Hence the $U(N)$ subgroup of this $U(2N)$ symmetry which will act on $D_a$ as $D_a{\longrightarrow}g{D}_ag^{+}$ will also have to act on $L_a$ as $L_a{\longrightarrow}g{L}_ag^{+}$, i.e 
\begin{eqnarray}
D_a^g=g{D}_ag^{+}~,~L_a^g=g{L}_ag^{+}~,~g{\in}U(N).\label{ori1}
\end{eqnarray}
It is not difficult to see that the two sets of gauge transformations (\ref{ori}) and (\ref{ori1}) are identical if we are looking at the action (\ref{ac1}) since it only depends on $X_a=\alpha D_a$ and not on $L_a$. However for the gauge field $A_a$ there is certainly a difference between the two sets of transformations (\ref{ori}) and (\ref{ori1}). Under (\ref{ori}) we have $A_a{\longrightarrow}gA_ag^++g[L_a,g^+]$ whereas under (\ref{ori1}) we have   $A_a{\longrightarrow}gA_ag^+$. The actions as written in (\ref{ck}) and (\ref{ck10}) are invariant only in the first case. 

We want thus to modify the definition of the link variable $W$ so that we have (\ref{ori}) and not (\ref{ori1}). In other words under this new definition the fixed background $L_a$ will not rotate whereas the gauge field $A_a$ will transform correctly as  $A_a{\longrightarrow}gA_ag^++g[L_a,g^+]$. Towards this end we introduce another covariant derivative $D_a^{'}=L_a+A_a^{'}$ through the gauged idempotent ${\gamma}_{D^{'}}$ given by 
\begin{eqnarray}
&&{\gamma}_{D^{'}}=\hat{\gamma}^{'}\frac{1}{\sqrt{\hat{\gamma}^{'2}}}\nonumber\\
&&\hat{\gamma}^{'}=\frac{1}{N}(1+2{\sigma}_aD_a^{'})={\gamma}+\frac{2}{N}{\sigma}_aA_a^{'}~,~~\hat{\gamma}^{'2}=1+\frac{8\sqrt{c_2}}{N^2}{\Phi}^{'}+\frac{2}{N^2}{\epsilon}_{abc}{\sigma}_cF_{ab}^{'}.\label{eq00}
\end{eqnarray}
As before we will also set ${\Phi}^{'}=0$. From the two idempotents ${\gamma}_{D}$ and ${\gamma}_{D^{'}}$ we construct the link variable $W$ as follows
\begin{eqnarray}
W={\gamma}_{D^{'}}{\gamma}_{D}.
\end{eqnarray}
The extended  $U(2N)$ symmetry will then act on $W$ as follows 
\begin{eqnarray}
W{\longrightarrow}VWV^{+}~,~V{\in}U(2N). 
\end{eqnarray}
This transformation property of $W$ can only be obtained if we  impose the following transformation properties ${\gamma}_{D^{'}}{\longrightarrow}V{\gamma}_{D^{'}}V^{+}$ and ${\gamma}_D{\longrightarrow}V{\gamma}_DV^{+}$ on ${\gamma}_{D^{'}}$ and ${\gamma}_D$ respectively. Hence the $U(N)$ subgroup of this $U(2N)$ symmetry which will act on $D_a$ as $D_a{\longrightarrow}g{D}_ag^{+}$ will also have to act on $D_a^{'}$ as $D_a^{'}{\longrightarrow}g{D}_a^{'}g^{+}$, i.e 
\begin{eqnarray}
D_a^g=g{D}_ag^{+}~,~D_a^{'g}=g{D}_a^{'}g^{+}~,~g{\in}U(N).
\end{eqnarray}
Under these transformations the gauge fields $A_a$ and $A_a^{'}$ transform as  $A_a{\longrightarrow}gA_ag^++g[L_a,g^+]$ and $A_a^{'}{\longrightarrow}gA_a^{'}g^++g[L_a,g^+]$ respectively like we want. 

Remark also that for every fixed configuration $A_a^{'}$ the link variable $W$ contains the same degrees of freedom contained in $ {\gamma}_{D}$. To see this we will go to the basis in which ${\gamma}_{D^{'}}$ is diagonal, viz
\begin{eqnarray}
{\gamma}_{D^{'}}=\left(\begin{array}{cc}
{\bf 1}_{N+1} & 0\\
0 & -{\bf 1}_{N-1}
\end{array}
\right). 
\end{eqnarray}
In this basis $ {\gamma}_{D}$ and $W$ will have the following generic forms
\begin{eqnarray}
{\gamma}_{D}=\left(\begin{array}{cc}
{W}_1 & {W}_{12}\\
 {W}_{12}^+& {W}_2
\end{array}
\right)~,~W=\left(\begin{array}{cc}
{W}_1 & {W}_{12}\\
 -{W}_{12}^+& -{W}_2
\end{array}
\right). \label{decomp}
\end{eqnarray}
 ${W}_1={W}_1^+$ is an $(N+1)\times (N+1)$ matrix, ${W}_2={W}_2^+$ is an $(N-1)\times (N-1)$ matrix and ${W}_{12}$ is an $(N+1)\times (N-1)$ matrix whereas the hermitian adjoint ${W}_{12}^+$ is an $(N-1)\times (N+1)$ matrix. Since ${\gamma}_D^2=1$ or equivalently $W^+W=1$ we must also have the conditions 
\begin{eqnarray}
&&{W}_1^+W_1+{W}_{12}{W}_{12}^+=1\nonumber\\
&&{W}_2^+W_2+{W}_{12}^+{W}_{12}=1\nonumber\\
&&{W}_1{W}_{12}+{W}_{12}{W}_2=0.\label{decomp1}
\end{eqnarray}
Knowing ${W}_{12}$ will determine completely the matrix $W$ (or equivalently ${\gamma}_D$ ) and hence we have $2(N+1)(N-1)=2N^2-2$ degrees of freedom which agrees with (\ref{dimen}).

\subsection{The coordinate transformation $(A_1,A_2,A_3){\longrightarrow}(W,\Phi)$}
The main idea is that we want to reparametrize  the gauge field on ${\bf S}^2_L$ in terms of the fuzzy link variable $W$ and the normal scalar field $\Phi$. In other words we want to replace the triplet $(A_1,A_2,A_3)$ with  $(W,\Phi)$. It is the link variable $W$ which contains the  degrees of freedom of the gauge field which are tangent to the sphere as is shown by the result (\ref{dimen}). Thus in summary we have the coordinate transformation
 \begin{eqnarray}
(A_1,A_2,A_3){\longrightarrow}(W,\Phi)
\end{eqnarray}
First we need  to show that  we have indeed the correct measure. Namely one must show that we have 
\begin{eqnarray}
\int dA_1dA_2dA_3=c_N\int dWd\Phi \label{meas}
\end{eqnarray}
where $c_N$ is some constant of proportionality which can only depend on $N$. In order to compute the measure we will compute the quantity $Tr_{2N}(dW)^+dW$ where $Tr_{2N}$ denotes the $2N\times 2N$ dimensional trace. For this exercise the scalar field $\Phi$ will not be assumed to be fixed whereas the other gauge configuration $A_a^{'}$ and its corresponding normal scalar field ${\Phi}^{'}$ are supposed to be some constant backgrounds. From the definition $W={\gamma}_{D^{'}} {\gamma}_D$ and equations (\ref{eq0}) and (\ref{eq00}) one can easily compute
\begin{eqnarray}
W=1+\frac{2}{N}\gamma {\sigma}_aA_a+\frac{2}{N}{\sigma}_aA_a^{'}\gamma-\frac{2}{N}\Phi -\frac{2}{N}{\Phi}^{'} +O(\frac{1}{N^2})
\end{eqnarray}
or equivalently 
\begin{eqnarray}
dW=\frac{2}{N}\gamma {\sigma}_adA_a-\frac{2}{N}d\Phi +O(\frac{1}{N^2}).
\end{eqnarray}
Hence a straightforward calculation yields the measure
\begin{eqnarray}
Tr_{2N}(dW)^+dW=\frac{8}{N^2}Tr(dA_a)^2+\frac{8}{N^2}Tr(d\Phi)^2-\frac{16}{N^3}Trd\Phi d(L_aA_a+A_aL_a)+O(\frac{1}{N^3}).
\end{eqnarray}
By using the identity $2\sqrt{c_2}\Phi=A_aL_a+L_aA_a+A_a^2$ we arrive at the result
\begin{eqnarray}
Tr_{2N}(dW)^+dW=\frac{8}{N^2}Tr(dA_a)^2-\frac{8}{N^2}Tr(d\Phi)^2+O(\frac{1}{N^3})
\end{eqnarray}
The correct ( more suggestive ) way of writing this equation is the following
\begin{eqnarray}
Tr(dA_a)^2=\frac{N^2}{8}Tr_{2N}(dW)^+dW+Tr(d\Phi)^2+O(\frac{1}{N}).
\end{eqnarray}
In the large $N{\longrightarrow}\infty $ limit it is obvious that this equation implies (\ref{meas}) which is what we desire. 

\subsection{The $U(1)$ gauge action as a linear one-plaquette model}
It remains now to show that the enlarged $U(2N)$ symmetry reduces to its $U(N)$ subgroup in the large $N$ limit. The starting point is the $2N-$dimensional one-plaquette action with a positive coupling constant ${\lambda}$, viz
\begin{eqnarray}
S_P^{}&=&\frac{N}{{\lambda}}Tr_{2N}(W+W^{+}-2)\label{plaquette1}
\end{eqnarray}
with the constraints
\begin{eqnarray}
W={\gamma}_{D^{'}}{\gamma}_{D}~,~{\Phi}=\frac{D_a^2-c_2}{2\sqrt{c_2}}.
\end{eqnarray}
We have the path integral

\begin{eqnarray}
Z_P=c_N^2\int d{\gamma}_{D^{'}}d{\Phi}^{'} {\delta}({\Phi}^{'})\int_{W={\gamma}_{D^{'}}{\gamma}_{D}}dWd\Phi {\delta}({\Phi})e^{S_P^{}}.\label{path0}
\end{eqnarray}
$c_N^2$ is the constant which appears in (\ref{meas}).  The extra integrations over ${\gamma}_{D^{'}}$ and ${\Phi}^{'}$ ( in other words over $D^{'}_a$ ) is included in order to maintain gauge invariance of the path integral. The integration over $W$ is done along the orbit $W={\gamma}_{D^{'}}{\gamma}_{D}$ inside the full $U(2N)$ gauge group. In above we have also to integrate over configurations $D_a$ and  $D^{'}_a$ such  that $\Phi=0$ and ${\Phi}^{'}=0$ since we are only interested in the limit $m{\longrightarrow}\infty $ of the model (\ref{ck10}). Furthermore we can conclude from the result (\ref{meas}) that in the large $N$ limit this path integral can be written as
\begin{eqnarray}
Z_P=\int dA^{'}_a{\delta}({\Phi}^{'})\int_{W={\gamma}_{D^{'}}{\gamma}_{D}}dA_a {\delta}({\Phi})e^{S_P}.\label{path}
\end{eqnarray}
We need now to check what happens to the action $S_P^{}$ in the large $N$ limit.  This is done in appendix $C$ and one finds

\begin{eqnarray}
S_P^{}&=&\frac{N}{{\lambda}}\bigg[-\frac{32}{N^2}Tr\bar{A}_a^2+\frac{16}{N^4}Tr\bigg(i[L_a,\bar{A}_b]-i[L_b,\bar{A}_a]+{\epsilon}_{abc}\bar{A}_c\bigg)^2+O(\frac{1}{N^5})\bigg].\label{151}
\end{eqnarray}
The constraints $D_a^2=c_2$ and $D_a^{'2}=c_2$ ( or equivalently ${\Phi}=0$ and ${\Phi}^{'}=0$ ) become in terms of the variables $\bar{A}_a=\frac{1}{2}A_a-\frac{1}{2}A_a^{`}$ and $S_a=D_a+D_a^{`}$ ( or equivalently  $\hat{A}_a=\frac{1}{2}A_a+\frac{1}{2}A_a^{`}$ ) 
\begin{eqnarray}
&&S_a^2+4\bar{A}_a^2=4c_2~ {\Leftrightarrow}~\hat{A}_a^2+\{L_a,\hat{A}_a\}+\bar{A}_a^2=0 \nonumber\\
&&\{S_a,\bar{A}_a\}=0~{\Leftrightarrow}~\{L_a,\bar{A}_a\}+\{\hat{A}_a,\bar{A}_a\}=0.
\end{eqnarray}
In the continuum limit these two constraints becomes $n_a\hat{A}_a=$ and $n_a\bar{A}_a=0$ respectively. By using the first constraint we can rewrite the action in the form
\begin{eqnarray}
S_P^{}&=&\frac{N}{{\lambda}}\bigg[\frac{64}{N^2}Tr\hat{A}_aL_a+\frac{32}{N^2}Tr\hat{A}_a^2+\frac{16}{N^4}Tr\bigg(i[L_a,\bar{A}_b]-i[L_b,\bar{A}_a]+{\epsilon}_{abc}\bar{A}_c\bigg)^2+O(\frac{1}{N^5})\bigg].\nonumber\\\label{153}
\end{eqnarray}
The leading contribution  in the action $S_P$ as written in equation (\ref{151}) is a simple Gaussian which is clearly dominated by the configuration $\bar{A}_a=0$. As a consequence the full path integration over $\bar{A}_a$ is dominated by $\bar{A}_a=0$. This yields a zero action which is obviously not what we want. Furthermore the path integration over $\hat{A}_a$ diverges since this action (\ref{151}) does not depend on these matrices. On the other hand the Gaussian term becomes in equation (\ref{153}) ( after using the constraint ) a quadratic  integral over the matrices  $\hat{A}_a$ but with a wrong sign since the first term converges to $0$ in the limit ( see appendix $B$ ). Thus the path integration over the three matrices $\hat{A}_a$ will again diverge. The one-plaquette action $S_P$ by itself is therefore not enough to obtain a $U(1)$ action on the sphere in the continuum large $N$ limit.

\subsection{A quadratic one-plaquette action }
Towards the end of constructing a $U(1)$ action on the fuzzy sphere using the one-plaquette variable $W$ we add to the action $S_P$ the following quadratic one-plaquette action  ( where ${\lambda}^{'}$ is the corresponding coupling constant )

\begin{eqnarray}
S_P^{'}&=&-\frac{N}{{\lambda}^{'}}Tr_{2N}(W^2+W^{+2}-2)\label{plaquette2}
\end{eqnarray}
Remark the extra minus sign in front of this action, i.e ${\lambda}^{'}$ is a positive coupling constant. As before we need now to compute the large $N$ limit of this quadratic one-plaquette action. This is also done in appendix $C$ and one finds the result
\begin{eqnarray}
S_P^{'}&=&-\frac{N}{{\lambda}^{'}}\bigg[\frac{256}{N^2}Tr\hat{A}_aL_a+\frac{128}{N^2}Tr\hat{A}_a^2+\frac{512}{N^4}Tr(\hat{A}_a^2+\{\hat{A}_a,L_a\})^2+\frac{64}{N^4}Tr\bigg(i[L_a,\bar{A}_b]\nonumber\\
&-&i[L_b,\bar{A}_a]+{\epsilon}_{abc}\bar{A}_c\bigg)^2+O(\frac{1}{N^5})\bigg].\label{172}
\end{eqnarray}

By putting the one-plaquette actions (\ref{153}) and (\ref{172}) together we obtain the total one-plaquette action
\begin{eqnarray}
S_P+S_P^{'}&=&-\frac{32}{{\lambda}_1N}Tr(\{\hat{A}_a,L_a\}+\hat{A}_a^2)-\frac{512}{{\lambda}^{'}N^3}Tr\big(\{\hat{A}_a,L_a\}+\hat{A}_a^2\big)^2-\frac{16}{{\lambda}_1N^3}Tr\bigg(i[L_a,\bar{A}_b]\nonumber\\
&-&i[L_b,\bar{A}_a]+{\epsilon}_{abc}\bar{A}_c\bigg)^2+O(\frac{1}{\lambda N^4})-O(\frac{1}{{\lambda}^{'} N^4}).
\end{eqnarray}
The positive coupling constant ${\lambda}_1$ is defined in terms of $\lambda$ abd ${\lambda}^{'}$ by 
\begin{eqnarray}
-\frac{1}{{\lambda}_1}=\frac{1}{{\lambda}}-\frac{4}{{\lambda}^{'}}.
\end{eqnarray}
The effect of the dominant terms ( the first two terms in the above action ) is now precisely what we want. The path integral over the three matrices $\hat{A}_a$ is given by
\begin{eqnarray}
{\delta}(\bar{A}_a)&=&2^{3N^2}\int d\hat{A}_a{\delta}(\Phi){\delta}({\Phi}^{'})~\exp\bigg\{-\frac{32}{{\lambda}_1N}Tr(\{\hat{A}_a,L_a\}+\hat{A}_a^2)-\frac{512}{{\lambda}^{'}N^3}Tr\big(\{\hat{A}_a,L_a\}+\hat{A}_a^2\big)^2 \bigg\}\nonumber\\
&=&2^{2N^2}\int d\hat{A}_a{\delta}\bigg(\frac{1}{2}\{x_a,\bar{A}_a\}+\frac{1}{2\sqrt{c_2}}\{\hat{A}_a,\bar{A}_a\}\bigg){\delta}\bigg(\frac{1}{2\sqrt{c}_2}\hat{A}_a^2+\frac{1}{2}\{x_a,\hat{A}_a\}+\frac{1}{2\sqrt{c_2}}\bar{A}_a^2\bigg)~\exp\bigg\{\rm same\bigg\}\nonumber\\
&{\simeq}&2^{2N^2}{\delta}\bigg(\frac{1}{2}\{x_a,\bar{A}_a\}\bigg)\int d\hat{A}_a{\delta}\bigg(\frac{1}{2}\{x_a,\hat{A}_a\}\bigg)~\exp\bigg\{-\frac{32}{{\lambda}_1N}Tr\hat{A}_a^2-\frac{512}{{\lambda}^{'}N^3}Tr(\hat{A}_a^2)^2 \bigg\}.
\end{eqnarray}
In the large $N$ limit the first term in the exponent dominates ( see below ) and as a consequence the path integral over the three matrices $\hat{A}_a$ becomes a simple Gaussian. Since the second constraint inside the integral has the effect of reducing the number of independent matrices $\hat{A}_a$ to just two we obtain the final result
\begin{eqnarray}
{\delta}(\bar{A}_a)&{\simeq}&2^{2N^2}{\delta}\bigg(\frac{1}{2}\{x_a,\bar{A}_a\}\bigg)\int d\hat{A}_a{\delta}\bigg(\frac{1}{2}\{x_a,\hat{A}_a\}\bigg)~\exp\bigg\{-\frac{32}{{\lambda}_1N}Tr\hat{A}_a^2\bigg\}\nonumber\\
&{\simeq}&2^{2N^2}{\delta}\bigg(\frac{1}{2}\{x_a,\bar{A}_a\}\bigg)\bigg(\frac{N\pi{\lambda}_1}{32}\bigg)^{N^2}.
\end{eqnarray}
Another ( more correct ) way of understanding this result is to note that this path integral is dominated in the large $N$ limit by the configurations $\hat{A}_a=0$.

The path integral of the one-plaquette model ( with an action $S_P+S_P^{'}$ ) becomes in the large $N$ limit  as follows 
\begin{eqnarray}
Z_P^{'}=\int d\bar{A}_a{\delta}\bigg(\frac{1}{2}\{x_a,\bar{A}_a\}\bigg)e^{S_P^{\rm eff}}\label{177}
\end{eqnarray}
where

\begin{eqnarray}
S_P^{\rm eff}&=&N^2\log(\frac{N\pi{\lambda}_1}{8})-\frac{16}{{\lambda}_1 N^3}Tr\bigg(i[L_a,\bar{A}_b]-i[L_b,\bar{A}_a]+{\epsilon}_{abc}\bar{A}_c\bigg)^2+O(\frac{1}{\lambda N^4})-O(\frac{1}{{\lambda}^{'} N^4}).\nonumber\\
\end{eqnarray}
Notice that this action is invariant not only under the trivial original gauge transformation law $\bar{A}_a{\longrightarrow}\bar{A}_a$ but also it is invariant under the non-trivial gauge transformation $\bar{A}_a{\longrightarrow}\bar{A}_a+g[L_a,g^+]$ where $g{\in}U(N)$. This emergent new gauge transformation of $\bar{A}_a$ is identical to the transformation property of a $U(1)$ gauge field on the sphere. Therefore the action $S_P^{\rm eff}$ given by the above equation  is essentially the same $U(1)$ action $-(S-S_0)$ given in equation (\ref{aclim})  provided we make the following identification 
\begin{eqnarray}
\frac{16}{ N^2{\lambda}_1}\equiv \frac{16}{N^2}(-\frac{1}{{\lambda}}+\frac{4}{{\lambda}^{'}})=\frac{1}{4g^2}\equiv \frac{\tilde{\alpha}^4}{4}\equiv \frac{\bar{\alpha}^4}{4N^2}\label{mmm}
\end{eqnarray}
between the $U(1)$ gauge coupling constant $g$ on the fuzzy sphere and the one-plaquette model coupling constant ${\lambda}_1$. The action becomes
\begin{eqnarray}
S_P^{\rm eff}&=&N^2\log(\frac{8\pi g^2}{N})-\frac{1}{4g^2}\int \frac{d{\Omega}}{4\pi}(i{\cal L}_a\bar{A}_b-i{\cal L}_b\bar{A}_a+{\epsilon}_{abc}\bar{A}_c)^2+O(\frac{1}{\lambda N^4})-O(\frac{1}{{\lambda}^{'} N^4}).\nonumber\\\label{aclim1}
\end{eqnarray}
Let us remark that in this large $N$ limit in which $g$ is kept fixed the one-plaquette coupling constant ${\lambda}_1 $ goes to zero. Hence the fuzzy sphere action with fixed coupling constant $g$ corresponds in this particular limit to the one-plaquette gauge field in the weak regime and agreement between the two is expected only for weak couplings ( large values of $\tilde{\alpha}$ ). To see this more clearly we notice that in terms of $\lambda$ and ${\lambda}^{'}$ the limit ${\lambda}_1=\frac{\lambda {\lambda}^{'}}{4\lambda -{\lambda}^{'}}{\longrightarrow}0$ is equivalent to the limit ${\lambda}{\longrightarrow}0$ for fixed ${\lambda}^{'}$ or vice versa, i.e to the limit ${\lambda}^{'}{\longrightarrow}0$ for fixed ${\lambda}$. Furthermore ${\lambda}_1$ going to $0$ is also equivalent to the limit when both $\lambda$ and ${\lambda}^{'}$ go to zero. Clearly all these possibilities correspond to the one-plaquette gauge field in the weak regime.

Finally we remark that the constant term in (\ref{aclim}) depends on the mass parameter $m$. Thus by comparing between the constant terms in 
(\ref{aclim}) and (\ref{aclim1}) we can determine $m^2$ as a function of $g^2$ ( or equivalently $\bar{\alpha}^4$ ) and $N$. We find $m^2=\frac{32{\pi}^3g^6}{N}=\frac{32{\pi}^3N^5}{\bar{\alpha}^{12}}$.
  
\subsection{The one-plaquette path integral}
Instead of (\ref{path0}) we will therefore consider in the remainder of this article the following ( corrected or generalized ) one-plaquette path integral

\begin{eqnarray}
Z_P^{'}=c_N^2\int d{\gamma}_{D^{'}}d{\Phi}^{'} {\delta}({\Phi}^{'})\int_{W={\gamma}_{D^{'}}{\gamma}_{D}}dWd\Phi {\delta}({\Phi})e^{S_P^{}+S_P^{'}}.\label{path1}
\end{eqnarray}
In analogy with (\ref{decomp}) we  decompose the $2N{\times}2N$ matrices ${\gamma}_{D^{'}}{\gamma}_{D}$ and $W$ as follows 
\begin{eqnarray}
{\gamma}_{D^{'}}{\gamma}_{D}=\left(\begin{array}{cc}
({\gamma}_{D^{'}}{\gamma}_{D})_1 & ({\gamma}_{D^{'}}{\gamma}_{D})_{12}\\
 -({\gamma}_{D^{'}}{\gamma}_{D})_{12}^+& -({\gamma}_{D^{'}}{\gamma}_{D})_2
\end{array}
\right)~,~ W=\left(\begin{array}{cc}
{W}_1 & {W}_{12}\\
 -{W}_{12}^+& -{W}_2
\end{array}
\right). 
\end{eqnarray}
 In particular ${W}_1={W}_1^+$ is an $(N+1)\times (N+1)$ matrix, ${W}_2={W}_2^+$ is an $(N-1)\times (N-1)$ matrix and ${W}_{12}$ is an $(N+1)\times (N-1)$ matrix whereas the hermitian adjoint ${W}_{12}^+$ is an $(N-1)\times (N+1)$ matrix. Since $W^+W=1$ we have the conditions 
\begin{eqnarray}
&&{W}_1^+W_1+{W}_{12}{W}_{12}^+=1~,~{W}_2^+W_2+{W}_{12}^+{W}_{12}=1~,~{W}_1{W}_{12}+{W}_{12}{W}_2=0.\label{decomp10}
\end{eqnarray}
First we observe that in this basis the metric becomes
\begin{eqnarray}
Tr_{2N}(dW)^+dW&=&Tr_{N+1}(dW_1)^+dW_1+Tr_{N-1}(dW_2)^+dW_2\nonumber\\
&+&Tr_{N+1}dW_{12}(dW_{12})^++Tr_{N-1}(dW_{12})^+dW_{12}.
\end{eqnarray}
Hence we can immediately conclude that the path integral over $W$ can be rewritten  ( by neglecting an overall proportionality factor ) as
\begin{eqnarray}
\int_{W={\gamma}_{D^{'}}{\gamma}_{D}}dW{\propto} \int_{W_{12}=({\gamma}_{D^{'}}{\gamma}_{D})_{12}}dW_{12}\int_{W_{12}^+=({\gamma}_{D^{'}}{\gamma}_{D})_{12}^+}dW_{12}^+\int_{W_1=({\gamma}_{D^{'}}{\gamma}_{D})_1}dW_1\int_{W_2=({\gamma}_{D^{'}}{\gamma}_{D})_2}dW_2.\nonumber\\
\end{eqnarray}
Furthermore we can show that in this basis the actions $S_P^{}$ and $S_P^{'}$ take the form
\begin{eqnarray}
S_P&=&\frac{N}{\lambda}Tr_{N+1}(W_1+W_1^+-2)+\frac{N}{\lambda}Tr_{N-1}(-W_2-W_2^+-2)
\end{eqnarray}
and
\begin{eqnarray}
S_P^{'}=&-&\frac{N}{{\lambda}^{'}}Tr_{N+1}(W_1^2+W_1^{+2}-2)-\frac{N}{{\lambda}^{'}}Tr_{N-1}(W_2^2+W_2^{+2}-2)\nonumber\\
&+&\frac{2N}{{\lambda}^{'}}Tr_{N+1}W_{12}W_{12}^{+}+\frac{2N}{{\lambda}^{'}}Tr_{N-1}W_{12}^{+}W_{12}.
\end{eqnarray}
Thus the off-diagonal matrices $W_{12}$ and $W_{12}^{+}$ ( as opposed to the diagonal matrices $W_1$ and $W_2$ ) appear only in the action $S_P^{'}$.

 Let us recall that since the integration over $W$ is done along the orbit $W={\gamma}_{D^{'}}{\gamma}_{D}$ inside $U(2N)$ and since in the large $N$ limit both ${\gamma}_{D^{'}}$ and ${\gamma}_{D}$ approach the usual chirality operator $\gamma =n_a{\sigma}_a$ we see that $W$ approaches the identity matrix in this limit. It is in this sense that $W$ yields in the continuum large $N$ limit a small one-plaquette model. 

Let us now explain how we will approximate the above path integral in the continuum large $N$ limit. From one hand we have the following limiting constraint $W={\gamma}_{D^{'}}{\gamma}_{D}{\longrightarrow}{\bf 1}_{2N}$\footnote{ Notice that $W={\gamma}_{D^{'}}{\gamma}_{D}$ goes to ${\bf 1}_{2N}$ independently of any basis.} which means that when $N{\longrightarrow}\infty $  we have the behaviour $W_1=({\gamma}_{D^{'}}{\gamma}_{D})_1{\longrightarrow}{\bf 1}_{N+1}$, $-W_2=-({\gamma}_{D^{'}}{\gamma}_{D})_2{\longrightarrow}{\bf 1}_{N-1}$ and $W_{12}=({\gamma}_{D^{'}}{\gamma}_{D})_{12}{\longrightarrow}0$. From the other hand since $W$ must be always a unitary matrix and since the off-diagonal parts $W_{12}$ and  $W_{12}^+$ tend to zero the matrices  $W_1$ and $-W_2$ become in this approximations  $(N+1)\times (N+1)$ and $(N-1)\times (N-1)$ unitary matrices respectively ( which are close to the identity ) in accordance with equations (\ref{decomp10}). Let us also stress the fact that the strict limits of $W_1$, $-W_2$ and $W_{12}$ are independent of  ${\gamma}_{D^{'}}$. For example the matrix $W_1$ goes always to the same limit ${\bf 1}_{N+1}$ for all matrices ${\gamma}_{D^{'}}$ . 

The main approximation which we will adopt in this article consists therefore in replacing  the constraint $W={\gamma}_{D^{'}}{\gamma}_{D}$ with the simpler constraint $W{\longrightarrow}{\bf 1}_{2N}$ by taking the diagonal parts $W_1$ and $-W_2$  to be {\it two arbitrary}, i.e independent of ${\gamma}_{D^{'}}$, unitary matrices which are very close to the identities ${\bf 1}_{N+1}$ and ${\bf 1}_{N-1}$ respectively while allowing the off-diagonal parts  $W_{12}$ and $W_{12}^{+}$ to go to zero. We observe that by including only $W_1$ and $-W_2$  in this approximation we are including  in the limit precisely the correct number of degrees of freedom tangent to the sphere, viz $2N^2$. Thus in this approximation the integrations over $\Phi$, ${\Phi}^{'}$ and ${\gamma}_{D^{'}}$ decouple while the integrations over $W_{12}$ and $W_{12}^+$ are dominated by $W_{12}=W_{12}^+=0$. There remains the two independent path integrals over $W_1$ and $-W_2$  which  are clearly equal  in the strict limit since the matrix dimension of $W_1$ approaches  the matrix dimension of $-W_2$ for large $N$. Thus the path integral $Z_P^{'}$ reduces   ( by neglecting also an  overall proportionality factor )  to

\begin{eqnarray}
Z_P^{'}&{\propto}&[Z_P(\lambda,{\lambda}^{'})]^2\label{188}
\end{eqnarray}
where
\begin{eqnarray}
Z_P(\lambda,{\lambda}^{'})&=&\int dW_1 \exp\bigg\{\frac{N}{\lambda}Tr(W_1+W_1^+-2)-\frac{N}{{\lambda}^{'}}Tr(W_1^2+W_1^{+2}-2)\bigg\}.\label{189}
\end{eqnarray}

The path integral of  a $2-$dimensional $U(N)$ gauge theory in the axial gauge $A_1=0$  on a lattice with volume $V$ and lattice spacing $a$ is given by $Z_P(\lambda,\infty)^{V/a^2}$ where $Z_P(\lambda,\infty)$ is the above partition function (\ref{189}) for ${\lambda}^{'}=\infty $, i.e the partition function of the one-plaquette model $S_p=\frac{N}{{\lambda}}Tr(W_1+W_1^{+}-2)$. Next we need to understand the effect of the addition of the quadratic  one-plaquette action $S_p^{'}=-\frac{N}{{\lambda}^{'}}Tr(W_1^2+W_1^{+2}-2)$. Formally the partition function $Z_P(\lambda,{\lambda}^{'})^{V/a^2}$ for any value of the coupling constant ${\lambda}^{'}$ can be obtained by expanding the model $S_P+S^{'}_P$ around ${\lambda}^{'}=\infty $. Thus it is not difficult to observe that the one-plaquette action  $S_p+S_p^{'}$ does also lead to  ( a more complicated ) $U(N)$ gauge theory in two dimensions. The $U(N)$ gauge coupling constant $g^2_1$ is simply given by
\begin{eqnarray}
\frac{1}{g^2_1}=Na^4\frac{1}{\lambda}.\label{mmmm}
\end{eqnarray}
Therefore we can see that  the partition function $Z_P^{'}$ of a $U(1)$ gauge field on the fuzzy sphere  is proportional to the partition function of  a {\it generalized} $2-$dimensional $U(N)$ gauge theory in the axial gauge  $A_1=0$  on a lattice with two plaquettes. This doubling of plaquettes is reminiscent of the usual doubling of points in Connes standard model. The $U(1)$ gauge coupling constant $g^2$ and the $U(N)$ gauge coupling constant $g^2_1$ are related ( from (\ref{mmm}) and (\ref{mmmm}) ) by the equation
\begin{eqnarray}
\frac{g^2}{g^2_1}=\frac{N^3a^4}{64}\bigg(\frac{4{\lambda}}{{\lambda}^{'}}-1\bigg).
\end{eqnarray}
It is quite natural to require the two coupling constants $g^2$ and $g^2_1$ to be equal which means we must choose the lattice spacing $a$ such that 
\begin{eqnarray}
a^4=\frac{64}{N^3}\frac{{\lambda}^{'}}{4\lambda -{\lambda}^{'}}.
\end{eqnarray} 
\subsection{Saddle point solution}

We are therefore interested in the $N-$dimensional one-plaquette model
\begin{eqnarray}
Z_P(\lambda,{\lambda}^{'})&=&\int dW exp\bigg(\frac{N}{\lambda}Tr(W+W^+-2)-\frac{N}{{\lambda}^{'}}Tr(W^2+W^{+2}-2)\bigg).\label{193}
\end{eqnarray}
Let us recall that $dW$ is the $U(N)$ Haar measure. We can immediately diagonalize the link variable $W$ by writing $W=TDT^{+}$ where $T$ is some $U(N)$ matrix and $D$ is diagonal with elements equal to the eigenvalues $exp(i{\theta}_i)$ of $W$. In other words $D_{ij}={\delta}_{ij}exp(i{\theta}_i)$. The integration over $T$ can be done trivially and one ends up with the path integral
\begin{eqnarray}
Z_P(\lambda,{\lambda}^{'})=\int {\prod}_{i=1}^{N}d{\theta}_i e^{NS_N}.\label{194}
\end{eqnarray}
The action $S_N$ contains besides the Wilson actions $\frac{1}{\lambda}Tr(W+W^{+}-2)=\frac{2}{\lambda}\sum_{i=1}^N cos{\theta}_i-\frac{2N}{\lambda}$ and $\frac{1}{{\lambda}^{'}}Tr(W^2+W^{+2}-2)=\frac{2}{{\lambda}^{'}}\sum_{i=1}^N cos2{\theta}_i-\frac{2N}{{\lambda}^{'}}$ contributions coming from the usual Vandermonde determinant. Explicitly the total action reads

\begin{eqnarray}
S_N&=& \frac{2}{\lambda}\sum_{i} \cos{\theta}_i-\frac{2}{{\lambda}^{'}}\sum_{i} \cos2{\theta}_i+\frac{1}{2N}\sum_{i{\neq}j}\ln \bigg(\rm sin\frac{{\theta}_i-{\theta}_j}{2}\bigg)^2-\frac{2N}{\lambda}+\frac{2N}{{\lambda}^{'}}.\label{1p}
\end{eqnarray}

In the large $N$ limit we can resort to the method of steepest descent to evaluate the path integral $Z_P(\lambda,{\lambda}^{'})$ . The partition function will be dominated by the solution of the equation $\frac{dS_N}{d{\theta}_i}=0$ which is a minimum of the action $S_N$. Before we proceed to the solution we need to take into account the following crucial property. Since the link variable $W$ tends to one in the large $N{\longrightarrow}\infty$ limit we can conclude that all the angles ${\theta}_i$ tend to $0$ in this limit and thus we can   consider instead of the full one-plaquette model action 
(\ref{1p}) a {\it small} one-plaquette model action by including corrections up to the quadratic order in the angles ${\theta}_i$. We obtain
\begin{eqnarray}
S_N&=&-\frac{1}{{\lambda}_2}\sum_i {\theta}_i^2+\frac{1}{2N}\sum_{i{\neq}j}\ln \frac{\big({\theta}_i-{\theta}_j\big)^2}{4}+O({\theta}^4).
\label{exx}
\end{eqnarray}
${\lambda}_2$ is given by 
\begin{eqnarray}
&&\frac{1}{{\lambda}_2}=-\frac{1}{{\lambda}_1}+\frac{1}{12}.
\end{eqnarray}
For the consistency of the solution below the coupling constant ${\lambda}_1$ must be negative ( as opposed to the classical model  where ${\lambda}_1$ was assumed positive ) and as a consequence the coupling constant ${\lambda}_2$ is always positive. As it turns out most of the classical arguments of sections $4.2$ and $4.3$ will go through unchanged when  ${\lambda}_1$  is taken negative.

Thus in the following quantum theory of the model we will identify the effective one-plaquette action $S_P^{\rm eff}$ with the fuzzy sphere action 
$S-S_0$ ( which is to be compared with the classical identification $-S_P^{\rm eff}=S-S_0$) and hence we must make the following identification of the coupling constants
\begin{eqnarray}
-\frac{16}{ N^2{\lambda}_1}= \frac{1}{4g^2}=\frac{\bar{\alpha}^4}{4N^2}.
\end{eqnarray}
This is precisely due  as we have said to the fact that ${\lambda}_1$ becomes negative in the quantum theory. In the continuum large $N$ limit where $ \tilde{\alpha}^4$ is kept fixed instead of ${\lambda}_1$ we can see that $\frac{1}{{\lambda}_1}$ scales with $N^2$ and as a consequence
\begin{eqnarray}
{\lambda}_2=-{{\lambda}_1}=\frac{64}{N^2\tilde{\alpha}^4}.\label{off}
\end{eqnarray}

The saddle point solution must satisfy the equation of motion
\begin{eqnarray}
\frac{2}{{\lambda}}\sin{\theta}_i-\frac{4}{{\lambda}^{'}}\sin2{\theta}_i=\frac{1}{N}\sum_{j{\neq}i}\cot\frac{{\theta}_i-{\theta}_j}{2}.\label{eomla}
\end{eqnarray}

The   equation of motion (\ref{eomla}) takes ( in the limit $N{\longrightarrow}\infty $ when all the angles tend to zero ) the form

\begin{eqnarray}
\frac{2{\theta}_i}{{\lambda}_2}=\frac{2}{N}\sum_{j{\neq}i}\frac{1}{{\theta}_i-{\theta}_j}
\label{erlm2}
\end{eqnarray}
In order to solve the above problem we introduce the potential $V(\{{\theta}_i\})$ defined through its first derivative $\frac{dV(\{{\theta}_i\})}{d{\theta}_i}{\equiv}V^{'}({\theta}_i)
=\frac{2{\theta}_i}{{\lambda}_2}$ and also the $N\times N$ matrix $M$ defined through its eigenvalues ${\theta}_i$, $i=1,...,N$. The trace ${\omega}(z)$ of the resolvent of $M$ is given by
\begin{eqnarray}
{\omega}(z)=\frac{1}{N}Tr\frac{1}{M-z}=\frac{1}{N}\sum_{i}\frac{1}{{\theta}_i-z}.
\end{eqnarray}
The condition (\ref{erlm2}) can then be rewritten as follows
\begin{eqnarray}
{\omega}^2(z)-\frac{1}{N}{\omega}^{'}(z)+V^{'}(z){\omega}(z)=-R(z)\equiv -\frac{1}{N}\sum_{i}\frac{V^{'}(z)-V^{'}({\theta}_i)}{z-{\theta}_i}.
\end{eqnarray}
In the large $N$ limit we can also introduce a density of eigenvalues ${\rho}(\theta)$ which is positive definite and normalized to one ; ${\rho}(\theta)>0$, $\int d{\theta} {\rho}(\theta) =1 $ [ $N{\rho}(\theta)$ is the number of eigenvalues  in the range $[\theta-{d\theta}/{2},\theta + {d\theta}/{2}]$ ].  Thus the sum will be replaced by $\sum_{i}=N\int d\theta {\rho}(\theta)$ and one obtain 
\begin{eqnarray}
{\omega}^2(z)+V^{'}(z){\omega}(z)=-R(z)\equiv -\int_{-{\theta}_*}^{{\theta}_*} d\theta {\rho}(\theta)\frac{V^{'}(z)-V^{'}({\theta})}{z-{\theta}}.\label{208}
\end{eqnarray}
The trace of the resolvant is now given by
\begin{eqnarray}
{\omega}(z)=\int_{-{\theta}_*}^{{\theta}_*} d\theta {\rho}(\theta)\frac{1}{{\theta}-z}.
\end{eqnarray}
The density of eigenvalues ${\rho}(\theta)$ should satisfy $\int_{-{\theta}_*}^{{\theta}_*} d\theta {\rho}(\theta)=1$ and ${\rho}(\theta){\geq}0$ for all angles $-{\theta}_* {\leq}\theta {\leq}{{\theta}_*}$. 
We can easily solve this problem since we can compute 
\begin{eqnarray}
&&R(z)=\frac{2}{{\lambda}_2}
\end{eqnarray}
and

\begin{eqnarray}
\sigma(z)&=&\frac{2i}{{\lambda}_2}\sqrt{2{\lambda}_2-z^2}.\label{214}
\end{eqnarray}

The solution of the equation of motion is immediately given by
\begin{eqnarray}
{\omega}_{\pm}(z)&=&-\frac{1}{2}V^{'}(z){\pm}\frac{1}{2}{\sigma}(z)\nonumber\\
&=&-\frac{z}{{\lambda}_2}\pm\frac{i}{{\lambda}_2}\sqrt{2{\lambda}_2 -z^2}.
\end{eqnarray}
The function ${\omega}(z)$ is a multi-valued function of $z$ with branch points at $z=\pm z_0=\pm\sqrt{2{\lambda}_2}$. Since the potential $V$ has only one minimum at $\theta =0$ the density of eigenvalues  must have only one support centered around this minimum. This support is clearly in the range between $-z_0$ and $+z_0$. In terms of the resolvent ${\omega}(z)$ the density of eigenvalues is defined by
\begin{eqnarray}
&&{\rho}(z)=\frac{{\omega}(z+i\epsilon)-{\omega}(z-i\epsilon)}{2\pi i}.
\end{eqnarray}
${\omega}(z+i\epsilon)$ is the trace of the resolvent of $M$ computed with a contour in the upper half complex plane and we choose for it the plus sign, viz ${\omega}(z+i\epsilon)={\omega}_{+}(z)$. Similarly ${\omega}(z-i\epsilon)$ is the trace of the resolvent of $M$ computed with a contour in the lower half complex plane and we choose for it the minus sign, viz ${\omega}(z-i\epsilon)={\omega}_{-}(z)$. We obtain therefore
\begin{eqnarray}
&&{\rho}(\theta)=\frac{1}{\pi{\lambda}_2}\sqrt{2{\lambda}_2-{\theta}^2}.\label{rho}
\end{eqnarray}
It is obvious that this density of eigenvalues is only defined for angles $\theta$ which are in the range $-\sqrt{2{\lambda}_2} {\leq}\theta {\leq}\sqrt{2{\lambda}_2}$. However the value of the critical angle ${\theta}_*$ should be determined from the normalization condition $\int_{-{\theta}_*}^{{\theta}_*} d\theta {\rho}(\theta)=1$. This condition yields the value
\begin{eqnarray}
{\theta}_*=\sqrt{2{\lambda}_2}.\label{cr}
\end{eqnarray}
\subsection{The one-plaquette phase transition}
It is quite obvious that the action (\ref{exx}) is an excellent approximation of   (\ref{1p}) for all angles ${\theta}_i$ in the range
\begin{eqnarray}
-\frac{1}{2}{\leq}{\theta}_i{\leq}\frac{1}{2}.\label{ra1}
\end{eqnarray}
The particular value $\frac{1}{2}$ comes from the fact that the expansion of the quadratic one-plaquette action $S_P^{'}$ will converge to the original expression only for small ${\theta}_i$  in the above range.  The expansions of the linear one-plaquette action $S_P$ and of the Vandermonde action will converge to the original expressions  for  ${\theta}_i$  in the range $-{1}{\leq}{\theta}_i{\leq}{1}$. 

The solution (\ref{rho}) with the critical angle (\ref{cr}) is then  valid only for very small values of the coupling constant ${\lambda}_2$.  Indeed it is only in this regime of small ${\lambda}_{2}$ where the fuzzy sphere action with fixed coupling constant $g$ is expected to correspond to the one-plaquette model as we have discussed previously. However in order to find the critical value of ${\lambda}_{2}$ we need to extend the solution (\ref{rho}) to higher values of ${\lambda}_{2}$. To this end we note that the action (\ref{exx}) can also be obtained from the effective one-plaquette model
\begin{eqnarray}
S_p^{\rm eff}&=& \frac{2}{{\lambda}_2^{\rm eff}}Tr(W_{\rm eff}+W_{\rm eff}^{+}-2)\nonumber\\
&=&\frac{2}{{\lambda}_2^{\rm eff}}\sum_{i} \cos{\theta}_i^{\rm eff}-\frac{2N}{{\lambda}_2^{\rm eff}}.\label{lrim}
\end{eqnarray}
For small ${\theta}_i^{\rm eff}$ in the range
\begin{eqnarray} 
-{1}{\leq}{\theta}_i^{\rm eff}{\leq}{1} \label{ra2}
\end{eqnarray}
The total effective one-plaquette action becomes
\begin{eqnarray}
S_N^{\rm eff}&=& 
-\frac{1}{{\lambda}_2^{\rm eff}}\sum_{i} ({\theta}_i^{{\rm eff}})^2+\frac{1}{2N}\sum_{i{\neq}j}\ln \frac{\big({\theta}_i^{\rm eff}-{\theta}_j^{\rm eff}\big)^2}{4}+O(({\theta}^{\rm eff})^4).\label{lj}
\end{eqnarray}
The action (\ref{lj}) must be  identical to the action (\ref{exx}) and hence we must have $({\theta}_i^{\rm eff})^2=\frac{{\lambda}_2^{\rm eff}}{{\lambda}_2}{\theta}_i^2$. From the two ranges (\ref{ra1}) and (\ref{ra2}) we conclude that ${\theta}_i^{\rm eff}=2{\theta}_i$ and ${\lambda}_2^{\rm eff}=4{\lambda}_2$.
 
The saddle point solution of the action (\ref{lrim}) must satisfy the equation of motion
\begin{eqnarray}
\frac{2}{{\lambda}_2^{\rm eff}}\sin{\theta}_i^{\rm eff}=\frac{1}{N}\sum_{j{\neq}i}\cot\frac{{\theta}_i^{\rm eff}-{\theta}_j^{\rm eff}}{2}.\label{eomla}
\end{eqnarray}
In the continuum  large $N$ limit this equation becomes

\begin{eqnarray}
\frac{2}{{\lambda}_2^{\rm eff}}\sin{\theta}_{\rm eff}=\int_{}^{}d{\tau}_{\rm eff} {\rho}({\tau}_{\rm eff})\cot\frac{{\theta}_{\rm eff}-{\tau}_{\rm eff}}{2}.
\end{eqnarray}
By using the expansion $\cot\frac{{\theta}-{\tau}}{2}=2\sum_{n=1}^{\infty}\big(\sin n\theta \cos n\tau -\cos n\theta \sin n\tau \big)$ we can solve this equation quite easily in the strong-coupling phase ( large values of ${\lambda}_2$ ) and one finds the solution
\begin{eqnarray}
{\rho}({\theta}_{\rm eff})=\frac{1}{2{\pi}}+\frac{1}{\pi {\lambda}_2^{\rm eff}} \cos {\theta}_{\rm eff}.
\end{eqnarray}
However it is obvious that this solution makes sense only  where the density of eigenvalues is positive definite, i.e for ${\lambda}_2^{\rm eff}$ such that
\begin{eqnarray}
\frac{1}{2\pi}-\frac{1}{\pi{\lambda}_2^{\rm eff}}{\geq}0~{\Leftrightarrow}({\lambda}_2^{\rm eff})^*=2~{\Leftrightarrow}{\lambda}_2^*=0.5.\label{199}
\end{eqnarray}
This strong-coupling solution should certainly work for large enough values of ${\lambda}_2$. However this is not the regime we want. To find the solution for small values of ${\lambda}_{2}$ the only difference  with the above analysis is that the range of the eigenvalues is now $[-{\theta}_*,+{\theta}_*]$ instead of $[-\pi,+\pi]$ where ${\theta}_*$ is an angle less than $\pi$ which is a function of ${\lambda}_{2}$. It is only in this regime of small ${\lambda}_{2}$ where the fuzzy sphere action with fixed coupling constant $g$ is expected to correspond to the one-plaquette. In the strong regime deviations become significant near the sphere-to-matrix transition. 
Finding the solution in the weak-coupling phase for the effective action (\ref{lj}) is a more involved exercise. This is done in \cite{gross} with the result
\begin{eqnarray}
{\rho}({\theta}_{\rm eff})=\frac{2}{{\pi}{\lambda}_2^{\rm eff}}\cos \frac{{\theta}_{\rm eff}}{2}\sqrt{\frac{{\lambda}_2^{\rm eff}}{2}-\sin^2\frac{{\theta}_{\rm eff}}{2}}.\label{199-1}
\end{eqnarray}
 \begin{eqnarray}
\sin\frac{{\theta}_{\rm eff}}{2}=\sqrt{\frac{{\lambda}_2^{\rm eff}}{2}}.
\end{eqnarray}
It is very easy to verify that the this density of eigenvalues and critical angle will reduce to the solution (\ref{rho}) and the critical angle (\ref{cr}) when the angles are taken to be very small.

The above computed  critical value ${{\lambda}_2}^*=0.5$ leads to the critical value of the coupling constant $\bar{\alpha}$
\begin{eqnarray}
\bar{\alpha}_*^4=\frac{64}{{\lambda}_2^*}=128~{\Leftrightarrow}~\bar{\alpha}_*=3.36
\end{eqnarray}
 which is to be compared with the observed value 
\begin{eqnarray}
\bar{\alpha}_*=3.35{\pm}0.25
\end{eqnarray}
Indeed for $U(1)$ theory we observe in Monte Carlo simulation of the model (\ref{ac1}) with the relation (\ref{1.15})  the value $\bar{\alpha}_*=3.35{\pm}0.25$ ( or equivalently the value ${{\lambda}_2}^*=0.51{\pm}0.15$ ). Indeed for very large values of the mass parameter $m$ we observe two critical lines ( see figure $2$); the lower line is the ${\bf S}^2_N$-to-matrix critical line discussed previously. This line comes from the measurement of the critical value ${\alpha}_s=\tilde{\alpha}_{*}$ from the action. The upper line is the one-plaquette critical line which we  can fit to the curve
 \begin{eqnarray}
{\alpha}_{p}=\big[\frac{0.04}{m^2}\big]^{\frac{1}{2}}+3.35{\pm}0.25. \label{pre3}
\end{eqnarray}
Remark that this curve saturates in the limit $m{\longrightarrow}{\infty}$ around the value $3.35$. The points ${\alpha}_p$   on figure $2$ comes  from the measurement of the position $\bar{\alpha}_{*}=\sqrt{N}\tilde{\alpha}_{*}$ of the peak in the specific heat which for large values of the mass captures the one-plaquette phase transition. For even larger values of $m$ the peak disappears and in this case ${\alpha}_{p}$ measures the position where the specific heat jumps discontinously to the value $1$. 
\begin{figure}[h]
\begin{center}
\includegraphics[width=7cm,angle=-90]{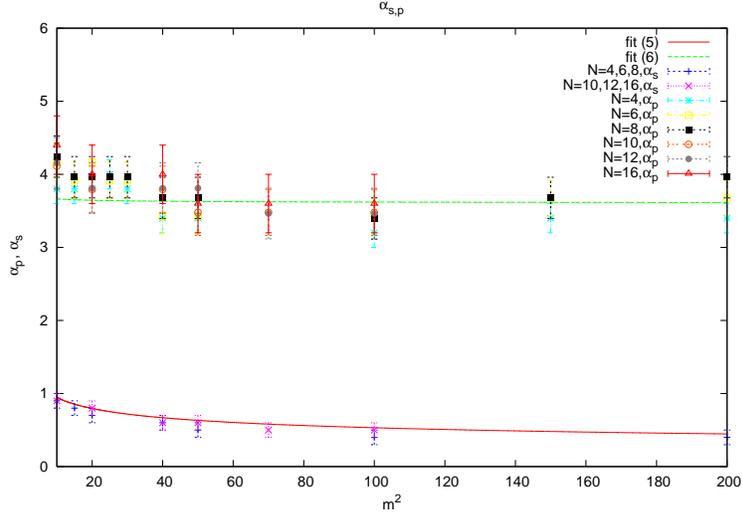}
\caption{{The phase diagram of the one-plaquette phase transition.}}
\end{center}
\end{figure}

\subsection{The specific heat and effective potential in $1/N$ expansion}
We are now in a position to compute the quadratic average $Q$ defined by the equation 

\begin{eqnarray}
Q=\frac{1}{N}\sum_{i}{\theta}_i^2.
\end{eqnarray}
We obtain in the fuzzy one-plaquette solution (\ref{rho}) the result
\begin{eqnarray}
Q=\int_{-{\theta}_*}^{{\theta}_*} d{\theta} {\rho}(\theta){\theta}^2&=&\frac{{\lambda}_2}{2}\label{224}
\end{eqnarray}
We need also to compute the non-local average

\begin{eqnarray}
Q_{\rm N-L}=\frac{1}{2N^2}\sum_{i{\neq}j}\ln\big(\frac{{\theta}_i-{\theta}_j}{2}\big)^2&=&\frac{1}{2}\int_{-{\theta}_*}^{{\theta}_*} d\theta {\rho}(\theta)\int_{-{\theta}_*}^{{\theta}_*}  d\alpha {\rho}(\alpha)\ln\big(\frac{{\theta}-{\alpha}}{2}\big)^2\nonumber\\
&=&\frac{1}{2}\ln \frac{{\lambda}_2}{2}+\frac{S_1}{2}.
\end{eqnarray}
$S_1$ is a constant of integration given explicitly by $S_1=\frac{4}{{\pi}^2}\int_{-1}^{1} dx \sqrt{1-x^2}\int_{-1}^{1} dy \sqrt{1-y^2}\ln\big(x-y\big)^2$.
The action (\ref{exx}) is therefore given by
\begin{eqnarray}
  \frac{S_N}{N}&=&-\frac{1}{{\lambda}_2}Q+Q_{\rm N-L}\nonumber\\
  &=&-\frac{1}{2}+\frac{1}{2}\ln\frac{{\lambda}_2}{2}+\frac{S_1}{2}.
\end{eqnarray} 
Let us recall from equations (\ref{188}) and (\ref{189}) that we have actually two identical one-plaquette models and hence the above action must be multiplied by a factor of $2$. Furthermore by comparing between  (\ref{177}) and (\ref{194}) we can see that  $S_P^{\rm eff}$ must be identified with $NS_N$ ( or twice as much due to the above factor  of $2$ ) whereas we have found that the  action $S-S_0$ on the fuzzy sphere must be identified in the quantum theory with $S_P^{\rm eff}$. In other words the effective action on the fuzzy sphere is given by

 \begin{eqnarray}
S&=&S_0+N^2(-1+\ln\frac{{\lambda}_2}{2}+S_1)\nonumber\\
&=&-\frac{1}{6}\tilde{\alpha}^4c_2-\frac{1}{2}\tilde{\alpha}^4c_2m^2-N^2\ln\tilde{\alpha}^4+N^2(-1-\ln\frac{N^2}{32}+S_1).\label{fit}
\end{eqnarray}
In above we have also used equation (\ref{off}). It is interesting to compare this effective action with the original effective action (\ref{emp})  obtained in the one-loop. If we set $\phi =1$ in (\ref{emp}) then we will find the same classical action as in the above equation, namely $-\frac{1}{6}\tilde{\alpha}^4c_2-\frac{1}{2}\tilde{\alpha}^4c_2m^2$. However the quantum correction in (\ref{emp}) in terms of $\tilde{\alpha}$ is by inspection given by $N^2\ln\tilde{\alpha}$ which is different from the quantum correction in the above equation which is equal to $-4N^2\ln\tilde{\alpha}$. We also note that in the large $m{\longrightarrow}\infty$, then large $N{\longrightarrow}\infty$ limit the above action will be dominated by the classical mass-dependent term  $-\frac{1}{2}\tilde{\alpha}^4c_2m^2$. This is precisely what we observe in Monte Carlo simulation. See figure $3$.
\begin{figure}[h]
\begin{center}
\includegraphics[width=7cm,angle=-90]{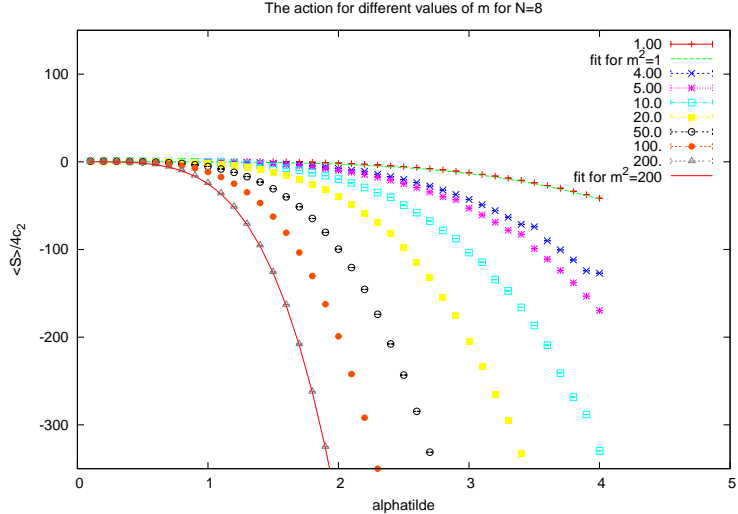}
\caption{{ The action for non-zero mass. The fit is given by the second term  of equation (\ref{fit}).}}
\end{center}
\end{figure}

Finally we need to compute the specific heat. Towards this end we implement the scaling transformations $S{\longrightarrow}\frac{S}{T}$ and $\tilde{\alpha}^4{\longrightarrow}\frac{\tilde{\alpha}^4}{T}$. The specific heat is then defined by
\begin{eqnarray}
C_{\rm v}=-\bigg(T^3\frac{{\partial}^2S}{{\partial}T^2}\bigg)_{T=1}.
\end{eqnarray}
A straightforward calculation yields the very simple result
\begin{eqnarray}
C_{\rm v}=N^2.
\end{eqnarray}
Again this is what we observe in our numerical simulation of the $U(1)$ gauge field on the fuzzy sphere in the weak regime. In the strong regime deviations are significant near the sphere-to-matrix transition. See figure $4$. In this regime of strong couplings the action and specific heat are computed using the distribution of eigenvalues (\ref{199-1}). We find

 \begin{eqnarray}
S&=&S_0+N^2(-\frac{1}{2}+\frac{1}{8{\lambda}_2^2}-\frac{1}{{\lambda}_2}+S_1)\nonumber\\
&=&-\frac{1}{6}\tilde{\alpha}^4c_2-\frac{1}{2}\tilde{\alpha}^4c_2m^2+\frac{N^2}{2}(\frac{\bar{\alpha}^4}{128})^2-N^2\frac{\bar{\alpha}^4}{64}+N^2(-\frac{1}{2}+S_1),
\end{eqnarray}
and
\begin{eqnarray}
C_{\rm v}=N^2(\frac{\bar{\alpha}^4}{128})^2.
\end{eqnarray}
We observe then that in the weak regime the specific heat is essentially given by $4c_2=N^2-1$ within statistical errors whereas in the strong regime the data does only follow the theoretical one-plaquette prediction away from the  ${\bf S}^2_N$-to-matrix transition. This is presumably due to the effects of the matrix phase which becomes strong near the critical ${\bf S}^2_N$-to-matrix transition. Remark that the minimum of the specific heat is where the ${\bf S}^2_N$-to-matrix transition happens for large values of $m$.

\begin{figure}[h]
\begin{center}
\includegraphics[width=7cm,angle=-90]{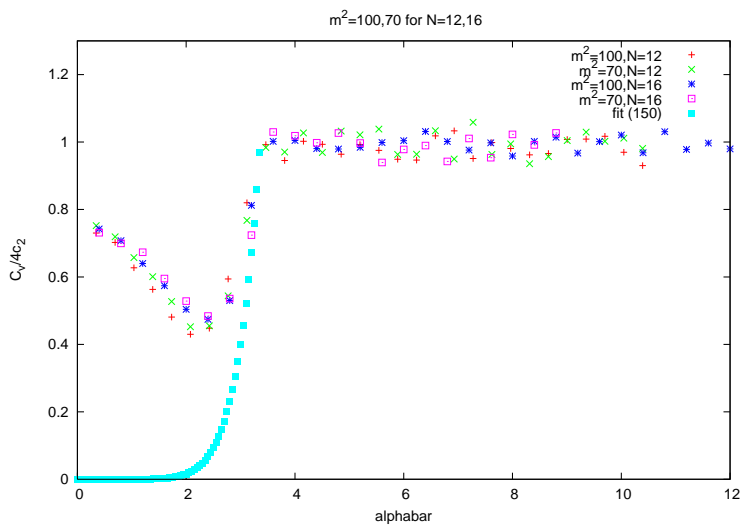}
\caption{{The specific heat for very large values of $m$.}}
\end{center}
\end{figure}

Let us comment further on the quantum effective potential for $U(1)$ gauge field on the fuzzy sphere
$S^2_{N}$ in this $1/N$ expansion of the one-plaquette model. As we have said before by comparing  the $1/N$ effective potential (\ref{fit}) with the one-loop effective potential (\ref{emp}) in which we set $\phi =1$ we can see that the classical contribution is the same in both potentials whereas the quantum correction in (\ref{emp}) in terms of $\tilde{\alpha}$ is given by $N^2\ln\tilde{\alpha}$ which is different from the quantum correction  $-4N^2\ln\tilde{\alpha}$ in (\ref{fit}). This observation allows us to rewrite ( or to guess that ) equation (\ref{fit}) ( should be rewritten ) in terms of the radius $\phi$ as follows
\begin{eqnarray}
S&=&\frac{N^2\tilde{\alpha}^4}{2}\bigg[\frac{1+m^2}{4}{\phi}^4-\frac{1}{3}{\phi}^3-\frac{m^2}{2}{\phi}^2\bigg]-4N^2\log\phi.
\end{eqnarray}
Now in contrast to what we have done so far in this article we will choose the mass parameter $m$ to be proportional to $N$. The simplest most natural choice is $m=\bar{m}N$. The effects of the large mass limit will then be included implicitly in the continuum large $N$ limit. This is in fact what was done in \cite{steinackers2}. The above effective potential becomes
\begin{eqnarray}
S&=&\frac{N^2\bar{\alpha}^4\bar{m}^2}{2}\bigg[\frac{1}{4}{\phi}^4-\frac{1}{2}{\phi}^2\bigg]-4N^2\log\phi.
\end{eqnarray}
It is very easy to verify that this potential admits a local minimum for all values of the coupling constant $\bar{\alpha}$. The minimum value ${\phi}_{\rm min}$ is found to be given by
\begin{eqnarray}
{\phi}_{\rm min}^2=\frac{1+\sqrt{1+\frac{32}{\bar{\alpha}^4\bar{m}^2}}}{2}.
\end{eqnarray}
In the limit ${m}{\longrightarrow}{\infty}$ first and then  ${N}{\longrightarrow}{\infty}$ ( considered in this article ) we can see that  $\bar{m}{\longrightarrow}{\infty}$ and hence ${\phi}_{\rm min}=1$ as expected. The fuzzy sphere ground state  (\ref{ac2}) is extremely stable in this limit and the the $1/N$ potential $S$ ( as opposed to the one-loop potential (\ref{emp}) ) is completely insensitive to the ${\bf S}^2_N-$to-matrix phase transition.

\section{Conclusion}
In this article we have shown explicitly that  quantum noncommutative $U(1)$ gauge field on the fuzzy sphere  ${\bf S}^2_N$ is equivalent ( at least in the fuzzy sphere-weak coupling phase )  to a quantum commutative $2-$dimensional $U(N)$ gauge field on a lattice with two plaquettes. 

By using the structure of the fuzzy sphere  we  have constructed a $2N{\times}2N$ matrix $W$ given by equation $(68)$ which was shown to contain the correct number of degrees of freedom tangent to the fuzzy sphere, namely $2N^2$ degrees of freddom. The other $N^2$ degrees of freedom are contained obviously in the normal scalar field $\Phi$. Indeed we have shown that  the gauge field $(A_1,A_2,A_3)$ is equivalent to $(W,\Phi)$ and $dA_1dA_2dA_3\propto dWd\Phi$. The fuzzy sphere action $(22)$ ( or equivalently $(13)$ ) written in terms of $A_a$ can therefore be rewritten in trems of $W$ and $\Phi$. We have shown explicitly that in the limit where we can set $\Phi=0$ this action $(22)$ and the action $S_P+S_P^{'}$ given by $(81)+(88)$ will tend to the same continuum limit, viz a $U(1)$ gauge theory on the ordinary sphere. As a consequence  the partition function of the fuzzy $U(1)$ model in the limit $m{\longrightarrow}\infty $ can be given either by the large $m$ limit of $(24)$ or  by equation $(98)$. 

Indeed the fuzzy partition function $Z_P^{'}$ given by equation $(98)$ is our starting point. It is found to be proportional to the partition function of a $U(N)$ model in the axial gauge $A_1=0$ on a lattice with two plaquettes given by equations  (\ref{188}) and (\ref{189}). We remark that the $U(N)$ theory consists of the canonical one-plaquette Wilson action $S_p=\frac{N}{{\lambda}}Tr(W_1+W_1^{+}-2)$ plus a novel quadratic  one-plaquette action $S_p^{'}=-\frac{N}{{\lambda}^{'}}Tr(W_1^2+W_1^{+2}-2)$ together with the canonical measure  $dW$. This is in fact the reason why $W$ is called a link variable. The quadratic term was needed in order that the fuzzy sphere one-plaquette path integral $Z_P^{'}$  converges to the sphere path integral in the large $N$ limit.  Remark also that the effective actions $S_p$ and $S_p^{'}$ involve the $N{\times}N$ link variable $W_1$ as opposed to the original actions $S_P$ and $S_P^{'}$ which involve the $2N{\times}2N$ link variable $W$. 

The doubling of plaquettes is a natural consequence of the model and it is reminiscent of the usual doubling of points in Connes standard model. However the ( other kind of ) doubling of $U(1)$ fuzzy gauge fields was needed in order to have a gauge invariant formulation of the fuzzy one-plaquette model. In fact a covariant plaquette variable $W$ can only be constructed out of two such $U(1)$ fuzzy fields. 

The main results are given by equations (\ref{188}) and (\ref{189}). It is therefore of paramount importance to find a more rigorous derivation of these two equations. Furthermore it will be very interesting to show that the large $m$ limit of the path integral $(24)$ and the path integral  $(98)$ are also equivalent for finite $N$. Their large $N$ equivalence used in this article  is confirmed in our Monte Carlo simulation of the model ( which uses $(24)$ ) where the measurement of the critical line $\bar{\alpha}_*=3.36$ and the specific heat can be understood in very simple terms using the limit $(105)$ of $(98)$. In particular the value $\bar{\alpha}_*=3.36$ seen in the simulation is precisely the Gross-Wadia-Witten one-plaquette $3$rd order transition point as calculated in this article from the path integral $(105)$.

Since the plaquette variable $W$ is small, i.e it approaches ${\bf 1}_{2N}$ in the large $N$ limit, we were able to show that the model in this limit reduces to a simple matrix model and as a consequence was easily solved. We computed the critical point and showed that it agrees with the observed value. We computed also the quantum effective potential and the specific heat  for $U(1)$ gauge field on the fuzzy sphere
$S^2_{N}$ in the $1/N$ expansion using this one-plaquette model. In particular the specific heat was found to be equal to $1$ in the fuzzy sphere-weak coupling phase of the gauge field which agrees with the observed value $1$ seen in Monte Carlo simulation. The value $1$ comes precisely because we have two plaquettes which approximate the noncommutative $U(1)$ gauge  field on the fuzzy sphere. In the fuzzy sphere-strong coupling phase deviations were found to be significant near the ${\bf S}^2_N-$to-matrix critical point. It will be very interesting to be able to extend this one-plaquette model to these large values of the gauge coupling constant $g^2$, i.e to small values of $\bar{\alpha}$. The key to this we believe lies in improving the basic approximations of this article  given in equations (\ref{188}) and (\ref{189}).

The most natural generalization of this work should include fermions in two dimensions \cite{15} and as a consequence take into account topological excitations \cite{14}. The best example which comes to mind is the Schwinger model \cite{16}. Then one should seriously contemplate going to $4$ dimensions with the full might of QCD. Early steps towards this larger goal were taken in the first reference of \cite{S2S2}. First we need to have a complete control over the phase diagram of the pure gauge model considered in this article \cite{ref1}.
\paragraph{Acknowledgements}
The author Badis Ydri  would like to thank Denjoe O'Connor, P.Castro-Villarreal and  R.Delgadillo-Blando for their
extensive discussions and critical comments while this research
was in progress.

\appendix

\section{Next-to-leading correction of the effective potential}
The next-to-leading contribution ( coming from the terms of order $N$ and of order $1$ in equations (\ref{1.34})-(\ref{1.36}) ) is given by
\begin{eqnarray}
Tr_2Tr_{N-1}\log\bigg[(1-\frac{2}{N}-\frac{4m^2}{N^2})\big({\delta}_{ij}-x_ix_j\big)+\frac{8m^2}{N^2}(1-\frac{1}{{\phi}^2}){\delta}_{ij}+\frac{4i}{N}(1-\frac{1}{N})(1-\frac{1}{\phi}){\epsilon}_{ij3}x_3\bigg].\nonumber\\
\end{eqnarray}
The ${\phi}-$dependence is only in the second and third terms inside the logarithm. Let us also remark that the inverse of the operator
\begin{eqnarray}
L_{ij}=a{\delta}_{ij}+bx_ix_j
\end{eqnarray}
is given by

\begin{eqnarray}
(L^{-1})_{ij}=\frac{1}{a}{\delta}_{ij}+\frac{1}{a}x_i\frac{1}{x_3^2-1-\frac{a}{b}}x_j.
\end{eqnarray}
If $a=-b=1-\frac{2}{N}-\frac{4m^2}{N^2}$ then we can see that the inverse $L^{-1}$ does not exist because of the zero eigenvalue of $x_3$. This can be traced to the fact that the rotational  $U(1)$ symmetry ( unlike gauge symmetry ) can not be restored back to the original full $SU(2)$ invariance.

Thus we regularize $L$ as follows
\begin{eqnarray}
&&L_{ij}=(1-\frac{2}{N}-\frac{4m^2}{N^2})\big({\epsilon}{\delta}_{ij}-x_ix_j\big).
\end{eqnarray}
Furthermore the value of $\phi$ in the matrix phase is expected to be very close to the classical value $1$. This means in particular that $1-\frac{1}{\phi}$ is a small number and thus it is a very good expansion parameter in this model. The vertex is given by

\begin{eqnarray}
V_{ij}=\frac{8m^2}{N}(1-\frac{1}{{\phi}^2}){\delta}_{ij}+4i(1-\frac{1}{N})(1-\frac{1}{\phi}){\epsilon}_{ij3}x_3.
\end{eqnarray}
This is small both in $\frac{1}{N}$ expansion and in $1-\frac{1}{\phi}$ expansion. We need to evaluate  
\begin{eqnarray}
Tr_2Tr_{N-1}\log\big[1+\frac{1}{N}L^{-1}V\big]=\sum_{r=1}^{\infty}\frac{(-1)^{r+1}}{r}\frac{1}{N^r}Tr_{N-1}(L^{-1}V)_{i_1i_2}(L^{-1}V)_{i_2i_3}....(L^{-1}V)_{i_ri_1}.
\end{eqnarray}
Since $1)$ $L^{-1}$ and $V$ are $(N-1){\times}(N-1)$ ordinary matrices, $2)$ the trace $\frac{1}{N-1}Tr_{N-1}$ becomes the ordinary integral in the limit and $3)$ the operators $x_a$ go over to the coordinates $n_a$ on the sphere we conclude that the only non-vanishing term in the $N{\longrightarrow}{\infty}$ limit is the order $r=1$ term in the above equation. We have
\begin{eqnarray}
Tr_2Tr_{N-1}\log\big[1+\frac{1}{N}L^{-1}V\big]&=&\frac{1}{N}Tr_{N-1}(L^{-1}V)_{i_1i_1}+...\nonumber\\
&=&-4i(1-\frac{1}{\phi}){\epsilon}_{ij3}\frac{1}{N}Tr_{N-1}x_i\frac{1}{x_3^2-1+\epsilon}x_jx_3+...\nonumber\\
&=&-\frac{16}{N^2}(1-\frac{1}{\phi})Tr_{N-1}\frac{x_3^2-\frac{1}{2}}{x_3^2-1+\epsilon}+....
\end{eqnarray}
This term is clearly going to zero in the limit. In particular
the contribution of the zero eigenvalue of $x_3$ is going to zero as $\frac{1}{N^2}\frac{1}{1-\epsilon}$. 

We can check that the quantum correction of the effective potential coming from the terms of order $O(\frac{1}{N})$ in equations (\ref{1.34})-(\ref{1.36}) are also going to zero in the limit. Hence the full quantum correction to the effective potential is given by (\ref{all}).

\section{The Star product on ${\bf S}^2_L$}
The coherent states on ${\bf S}^2$ are constructed as follows. Let us introduce the $2-$dimensional rank one projector $P_{\frac{1}{2}}=\frac{1}{2}+n_a\frac{{\sigma}_a}{2}$. The requirement $P^2_{\frac{1}{2}}=P_{\frac{1}{2}}$ implies the condition $\vec{n}^2=1$. At the north pole we have $\vec{n}_0=(0,0,1)$ and the projector becomes $P_{\frac{1}{2}}^0$ which projects onto  the state $|\vec{n}_0,\frac{1}{2}>={\small\left(\begin{array}{c}
1 \\
0
\end{array}
\right)}$. In other words $P_{\frac{1}{2}}^{0}=|\vec{n}_0,\frac{1}{2}><\vec{n}_0,\frac{1}{2}|$. A generic point $\vec{n}$ on ${\bf S}^2$ is obtained by the rotation  $g$ such that $\vec{n}=g\vec{n}_0$. The corresponding state is $|\vec{n},\frac{1}{2}>=g|\vec{n}_0,\frac{1}{2}>$ and the corresponding projector is precisely $P_{\frac{1}{2}}$ which can  also be rewritten as $P_{\frac{1}{2}}=|\vec{n},\frac{1}{2}><\vec{n},\frac{1}{2}|$. 

The irreducible representation $\frac{L}{2}$ of $SU(2)$ can be obtained from the symmetric product of $L$ copies of the fundamental representation $\frac{1}{2}$. The $\frac{L}{2}-$representation of the element $g\in SU(2)$ is given by the matrix $U^{(\frac{L}{2})}(g)$  defined by
\begin{eqnarray}
U^{(\frac{L}{2})}(g)=g{\otimes}_s...{\otimes}_s g~,~L~{\rm times}.
\end{eqnarray} 
The $(L+1)-$dimensional rank one projector $P_{\frac{L}{2}}$ which defines the $\frac{L}{2}-$coherent state $|\vec{n},\frac{L}{2}>$ is given  as the $L-$fold symmetric tensor product of the level $\frac{1}{2}$ projector $P_{\frac{1}{2}}$, viz
\begin{eqnarray}
P_{\frac{L}{2}}\equiv|\vec{n},\frac{L}{2}><\vec{n},\frac{L}{2}|=P_{\frac{1}{2}}{\otimes}_s...{\otimes}_s P_{\frac{1}{2}}~,~L~{\rm times}.
\end{eqnarray} 
The coherent state $|\vec{n},\frac{L}{2}>$ can also be constructed as $|\vec{n},\frac{L}{2}>=U^{(\frac{L}{2})}(g)|\vec{n}_0,\frac{L}{2}>$ where   $|\vec{n}_0,\frac{L}{2}>$ is the coherent state  defined by the projector $P_{\frac{L}{2}}^0 \equiv|\vec{n}_0,\frac{L}{2}><\vec{n}_0,\frac{L}{2}|=P_{\frac{1}{2}}^0{\otimes}_s...{\otimes}_s P_{\frac{1}{2}}^0$.

To any $N\times N$ matrix $\phi$ ( where $N=L+1$ ) we associate an ordinary function ${\phi}_L(\vec{n})$ on  ${\bf S}^2$ given by
\begin{eqnarray}
{\phi}_L(\vec{n})=<\vec{n},\frac{L}{2}|\phi |\vec{n},\frac{L}{2}>.
\end{eqnarray}
The product of two matrices ${\phi}{\psi}$ is mapped to the star product  ${\phi}_{L}*{\psi}_{L}(\vec{n})$ defined by
\begin{eqnarray}
{\phi}_L*{\psi}_{L}(\vec{n})=<\vec{n},\frac{L}{2}|\phi{\psi}  |\vec{n},\frac{L}{2}>.
\end{eqnarray}
We can show that 
\begin{eqnarray}
{\phi}_L*{\psi}_{L}(\vec{n})=\sum_{k=0}^L\frac{(L-k)!}{k!L!}K_{a_1b_1}...K_{a_kb_k}\frac{\partial}{{\partial}n_{a_1}}...\frac{\partial}{{\partial}n_{a_k}}{\phi}_L(\vec{n})\frac{\partial}{{\partial}n_{b_1}}...\frac{\partial}{{\partial}n_{b_k}}{\psi}_L(\vec{n})\label{starpro}
\end{eqnarray}
where
\begin{eqnarray}
K_{ab}={\delta}_{ab}-n_an_b+i{\epsilon}_{abc}n_c.
\end{eqnarray}
Using these coherent states we can compute
\begin{eqnarray}
&&<\vec{n},\frac{L}{2}|L_a |\vec{n},\frac{L}{2}>=\frac{L}{2}n_a\nonumber\\
&&<\vec{n},\frac{L}{2}|[L_a,\phi] |\vec{n},\frac{L}{2}>={\cal L}_a{\phi}_L(\vec{n})
\end{eqnarray}
where ${\cal L}_a=-i{\epsilon}_{abc}n_b{\partial}_c$ and
\begin{eqnarray}
\frac{1}{N}Tr{\phi}{\psi}=\int \frac{d\Omega}{4\pi}{\phi}_L*{\psi}_{L}(\vec{n}).
\end{eqnarray}
As another example we will compute $\frac{2}{N^2}TrL_aA_a$  which appears in the expansion (\ref{153}) of the one-plaquette action. We have immediately

\begin{eqnarray}
\frac{2}{N^2}TrL_aA_a&=&\frac{1}{N^2}Tr(L_aA_a+A_aL_a)\nonumber\\
&=&\frac{N-1}{2N}\int \frac{d\Omega}{4\pi}(n_a*(A_a)_L+(A_a)_L*n_a).
\end{eqnarray}
It must be clear that $(A_a)_L$ is the function which corresponds to the $N\times N$ matrix $A_a$. In this article since we are mostly working with the  $N\times N$ matrices $A_a$ it is more easier to denote the corresponding functions $(A_a)_L$ ( in the very few places which appear ) by the same symbol $A_a$ without fear of confusion . By using  the star product (\ref{starpro}) we obtain the result 
\begin{eqnarray}
\frac{2}{N^2}TrL_aA_a
&=&\int \frac{d\Omega}{4\pi}({\Phi}_{\infty} +\frac{1}{N}{\partial}_aA_a-\frac{1}{N}n_a{\partial}_a{\Phi}_{\infty})\nonumber\\
&=&\int \frac{d\Omega}{4\pi}\bigg((1+\frac{3}{N}){\Phi}_{\infty} +\frac{1}{N}{\partial}_a(A_a-n_a{\Phi}_{\infty})\bigg).
\end{eqnarray}
This is an exact formula where ${\Phi}_{\infty}$ is defined by ${\Phi}_{\infty}=n_aA_a$. In the limit ${\Phi}_{\infty}$ becomes  exactly the normal component of $A_a$ and therefore $A_a-n_a{\Phi}_{\infty} $ is precisely the tangent gauge field on the sphere. Hence we can see directly that $\int d\Omega {\partial}_a(A_a-n_a\Phi)=0$. Furthermore since ${\Phi}_{\infty}$ is a constant equal to $0$ in the limit we can conclude that we have the final result 
\begin{eqnarray}
\frac{2}{N^2}TrL_aA_a=0.
\end{eqnarray}

\section{The continuum limits of the one-plaquette actions $S_P$ and $S_P^{'}$}
We need to check what happens to the action $S_P^{}$ in the large $N$ limit. we have

\begin{eqnarray}
S_P^{}&=&\frac{N}{{\lambda}}Tr_{2}(W+W^{+}-2).
\end{eqnarray}
We will introduce the covariant matrices $\bar{A}_a$ and $S_a$ defined  respectively by $2\bar{A}_a=D_a-D_a^{'}=A_a-A_a^{'}$ and $S_a=D_a+D_a^{'}=2(L_a+\hat{A}_a)$ where $\hat{A}_a$ is a gauge field defined by the matrices $2\hat{A}_a=A_a+A_a^{'}$. The measure becomes therefore $dA_a^{'}dA_a=2^{3N^2}d\bar{A}_ad\hat{A}_a$. We start with the expansion
\begin{eqnarray}
{\gamma}_{D}&=&\bigg(\hat{\gamma} -\frac{1}{N^2}\hat{\gamma} ({\epsilon}_{abc}{\sigma}_cF_{ab})+\frac{3}{2N^4}\hat{\gamma} ({\epsilon}_{abc}{\sigma}_cF_{ab})^2+O(\frac{1}{N^6})\bigg)
\end{eqnarray}
and a similar expansion for ${\gamma}_{D^{'}}$. We can now compute the first non-vanishing covariant terms in $Tr_{2N}W$ to be
\begin{eqnarray}
Tr_{2N}W=Tr_{2N}\bigg(\hat{\gamma}\hat{\gamma}^{'}&-&\frac{1}{N^2}\hat{\gamma}\hat{\gamma}^{'}({\epsilon}_{abc}{\sigma}_cF_{ab}^{'})-\frac{1}{N^2}\hat{\gamma}^{'}\hat{\gamma}({\epsilon}_{abc}{\sigma}_cF_{ab})\nonumber\\
&+&\frac{3}{2N^4}\hat{\gamma}^{'}\hat{\gamma}({\epsilon}_{abc}{\sigma}_cF_{ab})^2+\frac{3}{2N^4}\hat{\gamma}\hat{\gamma}^{'}({\epsilon}_{abc}{\sigma}_cF_{ab}^{'})^2\nonumber\\
&+&\frac{1}{N^4}\hat{\gamma}({\epsilon}_{abc}{\sigma}_cF_{ab})\hat{\gamma}^{'}({\epsilon}_{abc}{\sigma}_cF_{ab}^{'})+O(\frac{1}{N^6})\bigg).
\end{eqnarray}
Explicitly we have ( by reducing the $2N-$dimensional trace $Tr_{2N}$ to the $N-$dimensional trace $Tr$ ) the following first contribution
\begin{eqnarray}
Tr_{2N}\hat{\gamma}\hat{\gamma}^{'}+~{\rm h.c}&=&\frac{2}{N^2}Tr\big(1+4D_aD_a^{'}\big)+~{\rm h.c}\nonumber\\
&=&4N-\frac{32}{N^2}Tr\bar{A}_a^2\label{p11}
\end{eqnarray}
Next we have 
\begin{eqnarray}
Tr_{2N}\bigg(-\frac{1}{N^2}\hat{\gamma}\hat{\gamma}^{'}({\epsilon}_{abc}{\sigma}_cF_{ab}^{'})\bigg)+~{\rm h.c}&=&-\frac{4}{N^4}Tr\bigg({\epsilon}_{abc}D_cF_{ab}^{'}+{\epsilon}_{abc}D_c^{'}F_{ab}^{'}+4iD_aD_b^{'}F_{ab}^{'}\bigg)+~{\rm h.c}\nonumber\\
&=&-\frac{8}{N^4}Tr\bigg({\epsilon}_{abc}D_cF_{ab}^{'}+{\epsilon}_{abc}D_c^{'}F_{ab}^{'}+\big(i[D_a,D_b^{'}]-i[D_b,D_a^{'}]\big)F_{ab}^{'}\bigg)\nonumber\\
&=&-\frac{8}{N^4}Tr{\cal F}_{ab}F_{ab}^{'}
.\label{p22}
\end{eqnarray}
In above the matrices ${\cal F}_{ab}$ are defined by ${\cal F}_{ab}=F_{ab}+F_{ab}^{'}-4i[\bar{A}_a,\bar{A}_b]$. Similarly we can obtain
\begin{eqnarray}
Tr_{2N}\bigg(-\frac{1}{N^2}\hat{\gamma}^{'}\hat{\gamma}({\epsilon}_{abc}{\sigma}_cF_{ab}\bigg)+~{\rm h.c}
&=&-\frac{8}{N^4}Tr{\cal F}_{ab}F_{ab}.\label{p33}
\end{eqnarray}
Finally we need to evaluate the following three terms

\begin{eqnarray}
{\delta}S_P&=&Tr_{2N}\bigg(
\frac{3}{2N^4}\hat{\gamma}^{'}\hat{\gamma}({\epsilon}_{abc}{\sigma}_cF_{ab})^2+\frac{3}{2N^4}\hat{\gamma}\hat{\gamma}^{'}({\epsilon}_{abc}{\sigma}_cF_{ab}^{'})^2+\frac{1}{N^4}\hat{\gamma}({\epsilon}_{abc}{\sigma}_cF_{ab})\hat{\gamma}^{'}({\epsilon}_{abc}{\sigma}_cF_{ab}^{'})\bigg)\nonumber\\
&=&Tr_{2N}\bigg(
\frac{3}{2N^4}\hat{\gamma}^{'}\hat{\gamma}({\epsilon}_{abc}{\sigma}_cF_{ab})^2+\frac{3}{2N^4}\hat{\gamma}\hat{\gamma}^{'}({\epsilon}_{abc}{\sigma}_cF_{ab}^{'})^2+\frac{1}{N^4}({\epsilon}_{abc}{\sigma}_cF_{ab})\hat{\gamma}\hat{\gamma}^{'}({\epsilon}_{abc}{\sigma}_cF_{ab}^{'})\nonumber\\
&+&\frac{1}{N^4}[\hat{\gamma},{\epsilon}_{abc}{\sigma}_cF_{ab}]\hat{\gamma}^{'}({\epsilon}_{abc}{\sigma}_cF_{ab}^{'})\bigg).
\end{eqnarray}
We start by computing the last piece. To this end  we use the identity
\begin{eqnarray}
[\hat{\gamma},{\epsilon}_{abc}{\sigma}_cF_{ab}]&=&\frac{4i}{N}{\sigma}_a\{D_b,F_{ab}\}+\frac{2}{N}{\epsilon}_{abc}[D_c,F_{ab}]=2i{\sigma}_a\{x_b^D,F_{ab}\}+O(\frac{1}{N}).\label{lll}
\end{eqnarray}
``$O(\frac{1}{N})$'' stands for all other subleading terms which will yield corrections of the order of $\frac{1}{N^5}$ or higher to the action. The operators $x_a^D$  are {\it covariant} coordinates on the fuzzy sphere defined by $x_a^D=D_a/\sqrt{{c}_2}$. It is clear that in the large $N{\longrightarrow}\infty $ limit $x_a^D {\longrightarrow}n_a$ which are the usual coordinates on the ordinary sphere. Thus the only difference  between $x_a^D$  and the usual coordinates $x_a=L_a/\sqrt{c_2}$ on the fuzzy sphere is that under $U(N)$ gauge transformations we have $x_a^D {\longrightarrow}gx_a^Dg^+  $ as opposed to $x_a$ which remain fixed. However since ${\Phi}=0$ the operator $\{x_b^D,F_{ab}\}$ tends in the continuum limit to $2n_bF_{ab}$ which  vanishes identically. Hence $[\hat{\gamma},{\epsilon}_{abc}{\sigma}_cF_{ab}]=O(\frac{1}{N})$ and thus we obtain

\begin{eqnarray}
{\delta}S_P&=&Tr_{2N}\bigg(
\frac{3}{2N^4}\hat{\gamma}^{'}\hat{\gamma}({\epsilon}_{abc}{\sigma}_cF_{ab})^2+\frac{3}{2N^4}\hat{\gamma}\hat{\gamma}^{'}({\epsilon}_{abc}{\sigma}_cF_{ab}^{'})^2+\frac{1}{N^4}({\epsilon}_{abc}{\sigma}_cF_{ab})\hat{\gamma}\hat{\gamma}^{'}({\epsilon}_{abc}{\sigma}_cF_{ab}^{'})\nonumber\\&+&O(\frac{1}{N^5})\bigg).\label{eqa}
\end{eqnarray}
To evaluate the other terms we use the following remarkable identity
\begin{eqnarray}
\hat{\gamma}^{'}\hat{\gamma}&=&\frac{4}{N^2}(D_a^{'}D_a+\frac{1}{4})+\frac{1}{N^2}{\epsilon}_{abc}{\sigma}_c\bigg({\cal F}_{ab}+i\{D_a,\bar{A}_b\}-i\{D_b,\bar{A}_a\}+i\{D_a^{'},\bar{A}_b\}-i\{D_b^{'},\bar{A}_a\}\bigg)\nonumber\\
&=&\frac{4}{N^2}\bigg(\frac{N^2}{4}-2\bar{A}_a^2-\frac{1}{2}[\bar{A}_a,S_a]\bigg)+\frac{1}{N^2}{\epsilon}_{abc}{\sigma}_c\bigg({\cal F}_{ab}+i\{S_a,\bar{A}_b\}-i\{S_b,\bar{A}_a\}\bigg).\label{143}
\end{eqnarray}
or equivalently
\begin{eqnarray}
[\hat{\gamma},\hat{\gamma}^{'}]&=&\frac{4i}{N^2}{\epsilon}_{abc}{\sigma}_c\{D_a,D_b^{'}\}+\frac{4}{N^2}[D_a,D_a^{'}].\label{lll1}
\end{eqnarray}
and
\begin{eqnarray}
\{\hat{\gamma},\hat{\gamma}^{'}\}&=&2-\frac{16}{N^2}\bar{A}_a^2+\frac{2}{N^2}{\epsilon}_{abc}{\sigma}_c{\cal F}_{ab}.\label{lll2}
\end{eqnarray}
We see immediately that since we are already at order $\frac{1}{N^4}$ we can set in equation (\ref{eqa}) the following $\hat{\gamma}\hat{\gamma}^{'}{\simeq}1$ and $\hat{\gamma}^{'}\hat{\gamma}{\simeq}1$. Thus we  obtain
\begin{eqnarray}
{\delta}S_P+{\rm h.c}&=&Tr_{2N}\bigg(
\frac{3}{N^4}({\epsilon}_{abc}{\sigma}_cF_{ab})^2+\frac{3}{N^4}({\epsilon}_{abc}{\sigma}_cF_{ab}^{'})^2+\frac{2}{N^4}({\epsilon}_{abc}{\sigma}_cF_{ab})({\epsilon}_{abc}{\sigma}_cF_{ab}^{'})+O(\frac{1}{N^5})\bigg)\nonumber\\
&=&\frac{4}{N^4}Tr\bigg(3F_{ab}^2+3F_{ab}^{'2}+2F_{ab}F_{ab}^{'}\bigg)+O(\frac{1}{N^5}).\label{p44}
\end{eqnarray} 
The one-plaquette action $S_P^{}$ becomes ( by putting the contributions (\ref{p11}), (\ref{p22}) ,(\ref{p33}) and (\ref{p44}) together )

\begin{eqnarray}
S_P^{}&=&\frac{N}{{\lambda}}\bigg[-\frac{32}{N^2}Tr\bar{A}_a^2-\frac{8}{N^4}Tr{\cal F}_{ab}\bigg(F_{ab}+F_{ab}^{'}\bigg)+\frac{4}{N^4}Tr\bigg(3F_{ab}^2+3F_{ab}^{'2}+2F_{ab}F_{ab}^{'}\bigg)+O(\frac{1}{N^5})\bigg]\nonumber\\
\end{eqnarray}
We remark that $-2Tr{\cal F}_{ab}\big(F_{ab}+F_{ab}^{'}\big)+Tr\big(3F_{ab}^2+3F_{ab}^{'2}+2F_{ab}F_{ab}^{'}\big)=Tr(F_{ab}-F_{ab}^{'})^2+8iTr(F_{ab}+F_{ab}^{'})[\bar{A}_a,\bar{A}_b]$ and thus

\begin{eqnarray}
S_P^{}&=&\frac{N}{{\lambda}}\bigg[-\frac{32}{N^2}Tr\bar{A}_a^2+\frac{4}{N^4}Tr\big(F_{ab}-F_{ab}^{'}\big)^2+\frac{32i}{N^4}Tr(F_{ab}+F_{ab}^{'})[\bar{A}_a,\bar{A}_b]+O(\frac{1}{N^5})\bigg].\nonumber\\
\end{eqnarray}
By using the results $F_{ab}+F_{ab}^{'}=\frac{i}{2}[S_a,S_b]+2i[\bar{A}_a,\bar{A}_b]+{\epsilon}_{abc}S_c$ and $F_{ab}-F_{ab}^{'}=i[S_a,\bar{A}_b]+i[\bar{A}_a,S_b]+2{\epsilon}_{abc}\bar{A}_c$ we have 
\begin{eqnarray}
\frac{4}{N^4}Tr(F_{ab}-F_{ab}^{'})^2+\frac{32i}{N^4}Tr(F_{ab}+F_{ab}^{'})[\bar{A}_a,\bar{A}_b]&=&\frac{4}{N^4}Tr\bigg(2[S_a,\bar{A}_b][S_b,\bar{A}_a]-2[S_a,\bar{A}_b]^2+8\bar{A}_a^2\nonumber\\
&+&16i{\epsilon}_{abc}\bar{A}_c[S_a,\bar{A}_b]-4[S_a,S_b][\bar{A}_a,\bar{A}_b]-16[\bar{A}_a,\bar{A}_b]^2\bigg).\nonumber\\
\end{eqnarray}
We recall that $S_a=2L_a+2\hat{A}_a$ and that all commutators $[\bar{A}_a,\bar{A}_b],[\hat{A}_a,\bar{A}_b]$ and $[\hat{A}_a,\hat{A}_b]$ are of order $\frac{1}{N}$ and hence lead to terms of order $\frac{1}{N^5}$ in the action in the limit. With this approximation the transformation laws $\bar{A}_a{\longrightarrow}g\bar{A}_ag^+$ and $\hat{A}_a{\longrightarrow}g\hat{A}_ag^++g[L_a,g^+]$ become $\bar{A}_a{\longrightarrow}\bar{A}_a$ and $\hat{A}_a{\longrightarrow}\hat{A}_a+g[L_a,g^+]$ respectively. Thus we obtain 
\begin{eqnarray}
\frac{4}{N^4}Tr(F_{ab}-F_{ab}^{'})^2+\frac{32i}{N^4}Tr(F_{ab}+F_{ab}^{'})[\bar{A}_a,\bar{A}_b]&=&\frac{16}{N^4}Tr\bigg(2[L_a,\bar{A}_b][L_b,\bar{A}_a]-2[L_a,\bar{A}_b]^2+2\bar{A}_a^2\nonumber\\
&+&8i{\epsilon}_{abc}\bar{A}_c[L_a,\bar{A}_b]-4[L_a,L_b][\bar{A}_a,\bar{A}_b]\bigg)+O(\frac{1}{N^5}).\nonumber\\
\end{eqnarray}
The one-plaquette action takes therefore the form
\begin{eqnarray}
S_P^{}&=&\frac{N}{{\lambda}}\bigg[-\frac{32}{N^2}Tr\bar{A}_a^2+\frac{16}{N^4}Tr\bigg(i[L_a,\bar{A}_b]-i[L_b,\bar{A}_a]+{\epsilon}_{abc}\bar{A}_c\bigg)^2+O(\frac{1}{N^5})\bigg].
\end{eqnarray}

We find now the continuum limit of the quadratic action
\begin{eqnarray}
S_P^{'}&=&-\frac{N}{{\lambda}^{'}}Tr_{2N}(W^2+W^{+2}-2).
\end{eqnarray}
We have
\begin{eqnarray}
Tr_{2N}W^2&=&\bigg(Tr_{2N}(\hat{\gamma}^{'}\hat{\gamma})^2+\frac{2}{N^2}Tr_{2N}\hat{\gamma}^{'}\hat{\gamma}I_2+\frac{2}{N^4}Tr_{2N}\hat{\gamma}^{'}\hat{\gamma}I_4+\frac{1}{N^4}Tr_{2N}I_2^2+O(\frac{1}{N^6})\bigg)\nonumber\\
\end{eqnarray}
where
\begin{eqnarray}
&&I_2=-\hat{\gamma}^{'}\hat{\gamma}({\epsilon}_{abc}{\sigma}_cF_{ab})-\hat{\gamma}^{'}({\epsilon}_{abc}{\sigma}_cF_{ab}^{'})\hat{\gamma}
\end{eqnarray}
and
\begin{eqnarray}
I_4&=&\frac{3}{2}\hat{\gamma}^{'}\hat{\gamma}({\epsilon}_{abc}{\sigma}_cF_{ab})^2+\frac{3}{2}\hat{\gamma}^{'}({\epsilon}_{abc}{\sigma}_cF_{ab}^{'})^2\hat{\gamma}
+\hat{\gamma}^{'}({\epsilon}_{abc}{\sigma}_cF_{ab}^{'})\hat{\gamma}^{}({\epsilon}_{abc}{\sigma}_cF_{ab}^{}).
\end{eqnarray}
Straightforward computation using equation (\ref{143}) gives 
\bigskip
\begin{eqnarray}
Tr_{2N}(\hat{\gamma}^{'}\hat{\gamma})^2+{\rm h.c}&=&\frac{16}{N^4}Tr\big(\frac{N^2}{2}-4\bar{A}_a^2\big)^2+\frac{16}{N^4}Tr[\bar{A}_a,S_a]^2-\frac{8}{N^4}Tr\bigg(\{S_a,\bar{A}_b\}-\{S_b,\bar{A}_a\}\bigg)^2\nonumber\\
&+&\frac{8}{N^4}Tr{\cal F}_{ab}^2\nonumber\\
&=&4N+\frac{64}{N^4}Tr\bigg(4(\bar{A}_a^2)^2-N^2\bar{A}_a^2\bigg)+\frac{16}{N^4}Tr[\bar{A}_a,S_a]^2-\frac{8}{N^4}Tr\bigg(\{S_a,\bar{A}_b\}-\{S_b,\bar{A}_a\}\bigg)^2\nonumber\\
&+&\frac{8}{N^4}Tr{\cal F}_{ab}^2.
\end{eqnarray}
Explicitly we have 
\begin{eqnarray}
-\frac{8}{N^4}Tr\bigg(\{S_a,\bar{A}_b\}-\{S_b,\bar{A}_a\}\bigg)^2&=&-\frac{16}{N^4}Tr\bigg[[S_a,\bar{A}_b]^2-[S_a,\bar{A}_b][S_b,\bar{A}_a]-2[S_a,S_b][\bar{A}_a,\bar{A}_b]\nonumber\\
&+&4\bar{A}_a^2S_b^2-4\bar{A}_bS_bS_a\bar{A}_a\bigg]\nonumber\\
&=&-\frac{16}{N^4}Tr\bigg[[S_a,\bar{A}_b]^2-[S_a,\bar{A}_b][S_b,\bar{A}_a]-2[S_a,S_b][\bar{A}_a,\bar{A}_b]\nonumber\\
&-&4\bar{A}_a^2\bigg]+\frac{64}{N^4}Tr\bar{A}_bS_bS_a\bar{A}_a+\frac{64}{N^4}Tr\bigg(4(\bar{A}_a^2)^2-N^2\bar{A}_a^2\bigg).\nonumber\\
\end{eqnarray}
In the last line above we have used the constraint $S_a^2=4c_2-4\bar{A}_a^2$. By using the second constraint $S_a\bar{A}_a=-\bar{A}_aS_a$ we can rewrite this equation as
\begin{eqnarray}
-\frac{8}{N^4}Tr\bigg(\{S_a,\bar{A}_b\}-\{S_b,\bar{A}_a\}\bigg)^2&=&\frac{16}{N^4}Tr\bigg[-[S_a,\bar{A}_b]^2+[S_a,\bar{A}_b][S_b,\bar{A}_a]+2[S_a,S_b][\bar{A}_a,\bar{A}_b]\nonumber\\
&+&4\bar{A}_a^2\bigg]-\frac{16}{N^4}Tr[\bar{A}_a,S_a]^2+\frac{64}{N^4}Tr\bigg(4(\bar{A}_a^2)^2-N^2\bar{A}_a^2\bigg).
\end{eqnarray}
Thus we obtain the final exact expression
\begin{eqnarray}
Tr_{2N}(\hat{\gamma}^{'}\hat{\gamma})^2+{\rm h.c}&=&4N-\frac{128}{N^2}Tr\bar{A}_a^2+\frac{512}{N^4}Tr(\bar{A}_a^2)^2+\frac{8}{N^4}Tr{\cal F}_{ab}^2\nonumber\\
&+&\frac{16}{N^4}Tr\bigg[-[S_a,\bar{A}_b]^2+[S_a,\bar{A}_b][S_b,\bar{A}_a]+2[S_a,S_b][\bar{A}_a,\bar{A}_b]+4\bar{A}_a^2\bigg].\label{dd1}
\end{eqnarray}
The next computation is to find
\begin{eqnarray}
\frac{2}{N^2}Tr_{2N}\hat{\gamma}^{'}\hat{\gamma}I_2&=&-\frac{2}{N^2}Tr_{2N}\bigg((\hat{\gamma}^{'}\hat{\gamma})^2({\epsilon}_{abc}{\sigma}_cF_{ab})+(\hat{\gamma}\hat{\gamma}^{'})^2({\epsilon}_{abc}{\sigma}_cF_{ab}^{'})\bigg).\end{eqnarray}
We use equation (\ref{143}) in the form $\hat{\gamma}^{'}\hat{\gamma}=I+{\epsilon}_{abc}{\sigma}_cJ_{ab}=I+{\epsilon}_{abc}{\sigma}_c\big(\frac{1}{N^2}{\cal F}_{ab}+\frac{i}{N^2}K_{ab}\big)$.
The definition of the operators $I$, $J_{ab}=-J_{ba}$ and $K_{ab}=-K_{ba}$ is of course obvious. Thus we can compute
\begin{eqnarray}
-\frac{2}{N^2}Tr_{2N}(\hat{\gamma}^{'}\hat{\gamma})^2({\epsilon}_{abc}{\sigma}_cF_{ab})+{\rm h.c}&=&-\frac{4}{N^2}Tr{\epsilon}_{abc}({V}_c+{V}_c^+)F_{ab}.
\end{eqnarray}
The operator $V_c$ is defined in terms of $I$ and $J_{ab}$  as follows
\begin{eqnarray}
V_c=2i{\epsilon}_{abd}J_{ab}J_{cd} +{\epsilon}_{abc}(IJ_{ab}+J_{ab}I).
\end{eqnarray}
It is easy to check that the contribution of the first term $2i{\epsilon}_{abd}J_{ab}J_{cd}$ is of order $\frac{1}{N^5}$ at least 
whereas the contribution of the second term ${\epsilon}_{abc}(IJ_{ab}+J_{ab}I)$ is given by
\begin{eqnarray}
-\frac{32}{N^4}Tr{\cal F}_{ab}F_{ab}+O(\frac{1}{N^5})
\end{eqnarray}
The final result is 
\begin{eqnarray}
\frac{2}{N^2}Tr_{2N}\hat{\gamma}^{'}\hat{\gamma}I_2+{\rm h.c}&=&-\frac{32}{N^4}Tr{\cal F}_{ab}(F_{ab}+F_{ab}^{'})+O(\frac{1}{N^5}).\label{dd3}
\end{eqnarray}
Next we have to compute the following
\begin{eqnarray}
\frac{2}{N^4}Tr_{2N}\hat{\gamma}^{'}\hat{\gamma}I_4&=&\frac{2}{N^4}Tr_{2N}\bigg(\frac{3}{2}(\hat{\gamma}^{'}\hat{\gamma})^2({\epsilon}_{abc}{\sigma}_cF_{ab})^2+\frac{3}{2}(\hat{\gamma}\hat{\gamma}^{'})^2({\epsilon}_{abc}{\sigma}_cF_{ab}^{'})^2+(\hat{\gamma}\hat{\gamma}^{'})^2({\epsilon}_{abc}{\sigma}_cF_{ab}^{'})({\epsilon}_{abc}{\sigma}_cF_{ab})\nonumber\\
&+&\hat{\gamma}^{'}\hat{\gamma}\hat{\gamma}^{'}({\epsilon}_{abc}{\sigma}_cF_{ab}^{'})[\hat{\gamma},{\epsilon}_{abc}{\sigma}_cF_{ab}]\bigg).
\end{eqnarray}
Since $[\hat{\gamma},{\epsilon}_{abc}{\sigma}_cF_{ab}]$ is of order $\frac{1}{N}$ we obtain
\begin{eqnarray}
\frac{2}{N^4}Tr_{2N}\hat{\gamma}^{'}\hat{\gamma}I_4&=&\frac{2}{N^4}Tr_{2N}\bigg(\frac{3}{2}(\hat{\gamma}^{'}\hat{\gamma})^2({\epsilon}_{abc}{\sigma}_cF_{ab})^2+\frac{3}{2}(\hat{\gamma}\hat{\gamma}^{'})({\epsilon}_{abc}{\sigma}_cF_{ab}^{'})^2+(\hat{\gamma}\hat{\gamma}^{'})^2({\epsilon}_{abc}{\sigma}_cF_{ab}^{'})({\epsilon}_{abc}{\sigma}_cF_{ab})\bigg)\nonumber\\
&+&O(\frac{1}{N^5}).
\end{eqnarray}
In above we can also make the approximations $\hat{\gamma}\hat{\gamma}^{'},\hat{\gamma}^{'}\hat{\gamma}{\simeq}1$ since we are already at order $\frac{1}{N^4}$. Hence we obtain
\begin{eqnarray}
\frac{2}{N^4}Tr_{2N}\hat{\gamma}^{'}\hat{\gamma}I_4+{\rm h.c}&=&\frac{2}{N^4}Tr_{2N}\bigg(3({\epsilon}_{abc}{\sigma}_cF_{ab})^2+3({\epsilon}_{abc}{\sigma}_cF_{ab}^{'})^2+2({\epsilon}_{abc}{\sigma}_cF_{ab}^{'})({\epsilon}_{abc}{\sigma}_cF_{ab})\bigg)+O(\frac{1}{N^5})\nonumber\\
&=&\frac{8}{N^4}Tr_{2N}\bigg(3F_{ab}^2+3F_{ab}^{'2}+2F_{ab}F_{ab}^{'}\bigg)+O(\frac{1}{N^5}).\label{dd4}
\end{eqnarray}
Finally we need to compute 
\begin{eqnarray}
\frac{1}{N^4}Tr_{2N}I_2^2+{\rm h.c}&=&\frac{1}{N^4}Tr_{2N}\bigg(\hat{\gamma}^{'}\hat{\gamma}({\epsilon}_{abc}{\sigma}_cF_{ab})\hat{\gamma}^{'}\hat{\gamma}({\epsilon}_{abc}{\sigma}_cF_{ab})+\hat{\gamma}\hat{\gamma}^{'}({\epsilon}_{abc}{\sigma}_cF_{ab}^{'})\hat{\gamma}\hat{\gamma}^{'}({\epsilon}_{abc}{\sigma}_cF_{ab}^{'})\nonumber\\
&+&2(\hat{\gamma}^{'}\hat{\gamma})^2({\epsilon}_{abc}{\sigma}_cF_{ab})({\epsilon}_{abc}{\sigma}_cF_{ab}^{'})+2\hat{\gamma}\hat{\gamma}^{'}\hat{\gamma}({\epsilon}_{abc}{\sigma}_cF_{ab})[\hat{\gamma}^{'},{\epsilon}_{abc}{\sigma}_cF_{ab}^{'}]\bigg)+{\rm h.c}\nonumber\\
&=&\frac{2}{N^4}Tr_{2N}\bigg(({\epsilon}_{abc}{\sigma}_cF_{ab})^2+({\epsilon}_{abc}{\sigma}_cF_{ab}^{'})^2+2({\epsilon}_{abc}{\sigma}_cF_{ab})({\epsilon}_{abc}{\sigma}_cF_{ab}^{'})\bigg)+O(\frac{1}{N^5})\nonumber\\
&=&\frac{8}{N^4}Tr_{2N}\big(F_{ab}^2+F_{ab}^{'2}+2F_{ab}F_{ab}^{'}\big)+O(\frac{1}{N^5}).\label{dd5}
\end{eqnarray}
By putting equations (\ref{dd1}),(\ref{dd3}),(\ref{dd4}) and (\ref{dd5}) together the quadratic one-plaquette action becomes
\begin{eqnarray}
S_P^{'}&=&-\frac{N}{{\lambda}^{'}}\bigg[-\frac{128}{N^2}Tr\bar{A}_a^2+\frac{512}{N^4}Tr(\bar{A}_a^2)^2+\frac{16}{N^4}Tr\bigg(-[S_a,\bar{A}_b]^2+[S_a,\bar{A}_b][S_b,\bar{A}_a]+4\bar{A}_a^2\nonumber\\
&+&2[S_a,S_b][\bar{A}_a,\bar{A}_b]\bigg)+\frac{8}{N^4}Tr\big(F_{ab}-F_{ab}^{'})^2+\frac{64i}{N^4}Tr\big(F_{ab}+F_{ab}^{'})[\bar{A}_a,\bar{A}_b]-\frac{128}{N^4}Tr[\bar{A}_a,\bar{A}_b]^2\nonumber\\
&+&O(\frac{1}{N^5})\bigg].
\end{eqnarray}
As before if we drop all commutators $[\bar{A}_a,\bar{A}_b],[\hat{A}_a,\bar{A}_b]$ and $[\hat{A}_a,\hat{A}_b]$ ( since they are of order $\frac{1}{N}$ and hence lead to terms of order $\frac{1}{N^5}$ in the action ) then the limit of $S_P^{'}$ reduces to 
\begin{eqnarray}
S_P^{'}&=&-\frac{N}{{\lambda}^{'}}\bigg[\frac{256}{N^2}Tr\hat{A}_aL_a+\frac{128}{N^2}Tr\hat{A}_a^2+\frac{512}{N^4}Tr(\hat{A}_a^2+\{\hat{A}_a,L_a\})^2+\frac{64}{N^4}Tr\bigg(i[L_a,\bar{A}_b]\nonumber\\
&-&i[L_b,\bar{A}_a]+{\epsilon}_{abc}\bar{A}_c\bigg)^2+O(\frac{1}{N^5})\bigg].
\end{eqnarray}

\end{document}